\documentclass[12pt]{article}
\usepackage{a4,amsthm,amsfonts,latexsym,amssymb,
}
\font\notefont=cmsl8
\def\a{\alpha}

\def\c{\gamma}

\def\e{\varepsilon}
\def\s{\sigma}

\def\D{\Delta}

\def\p{\partial}
\def\x{{\hat{x}}}

\def\cG{{\cal G}}
\def\R{\mathbb R}

\newtheorem{thm}{Theorem}
\newtheorem{lemma}[thm]{Lemma}
\newtheorem{cor}[thm]{Corollary}
\newcommand{\bt}{\begin{thm}{\hspace{-.55em}\em{\bf {: }}}}
\newcommand{\et}{\end{thm}}
\newcommand{\bc}{\begin{cor}{\hspace{-.55em}\em{\bf {: }}}}
\newcommand{\ec}{\end{cor}}
\newcommand{\bl}{\begin{lemma}{\hspace{-.55em}\em{\bf {: }}}}
\newcommand{\el}{\end{lemma}}
\newcommand{\be}{\begin{equation}}
\newcommand{\ee}{\end{equation}}
\newcommand{\bea}{\begin{eqnarray}}
\newcommand{\eea}{\end{eqnarray}}
\newcommand{\beax}{\begin{eqnarray*}}
\newcommand{\eeax}{\end{eqnarray*}}

\newcommand{\Tr}{\mbox{\rm Tr}}
\newcommand{\mfr}[2]{{\textstyle\frac{#1}{#2}}}
\newcommand{\V}{V^{\rm TF}}

\begin{document}

\title{A NEW COHERENT STATES APPROACH TO SEMICLASSICS WHICH GIVES
  SCOTT'S CORRECTION
\thanks{Work partially supported by an EU TMR grant, by the Danish 
      research foundation center MaPhySto, and by a grant from the Danish research
      council.  \newline \hfill \copyright 2002 \ by the authors. This article may be
      reproduced in its entirety for non-commercial purposes. }
}

\author{
\begin{tabular}{ccc}
   Jan Philip
    Solovej
    & &Wolfgang L Spitzer\\
    \normalsize Department of Mathematics& &\normalsize Department of Mathematics\\ 
    \normalsize University of Copenhagen& &\normalsize University of California\\
    \normalsize Universitetsparken 5& &\normalsize Davis, One Shields Avenue\\
    \normalsize DK-2100 Copenhagen, Denmark& &\normalsize CA 95616-8633, USA\\
    \normalsize {\it e-mail\/}: solovej@math.ku.dk& &\normalsize {\it e-mail\/}: 
    spitzer@math.ucdavis.edu
  \end{tabular}
}
\date{Aug. 30, 2002}
\maketitle
\tableofcontents

\section{Introduction} 

There are various highly developed methods for establishing
semiclassical approximations. Probably the most refined method is
based on pseudo-differential and Fourier integral operator calculi. 
This extremely technical approach is well suited for getting good
or even sharp error estimates. Here, sharp refers to the optimal 
exponent of the semiclassical parameter in the error term. These sharp
estimates however often require strong regularity assumptions on the
operators being investigated. 

A different and very simple method based on coherent states gives
the leading order semiclassical asymptotics under optimal regularity assumptions.
    
The method of coherent states was used by Thirring~\cite{Thirring} and
Lieb~\cite{Lieb1} to give a very short and simple proof of the 
Thomas-Fermi energy asymptotics of large atoms and molecules; see also a 
recent improvement by Balodis Matesanz and Solovej~\cite{Matesanz-Solovej}. 
This asymptotics had been first proved by Lieb and
Simon in \cite{Lieb-Simon} using a Dirichlet-Neumann bracketing method. 

Because of the Coulomb singularity of the atomic potential the 
pseudo-differential techniques are not immediately applicable to the
Thomas-Fermi asymptotics. 
 
In fact, although the Coulomb singularity does not affect the leading
order Thomas-Fermi asymptotics, in the sense that it is purely
semi-classical, it does cause the first correction to be of a
non-semiclassical nature.
The first correction to the Thomas-Fermi asymptotics was 
predicted by Scott in \cite{Scott} and was later generalized to
molecules and formulated as a clear
mathematical conjecture in \cite{Lieb1}.
   
The first mathematical proof of the Scott correction for atoms
was given by Hughes~\cite{Hughes} 
(a lower bound) and by Siedentop and 
Weikard~\cite{Siedentop-Weikard} (both bounds) by WKB type methods.
Bach~\cite{Bach} proved the Scott correction for ions.

In \cite{Ivrii-Sigal}, Ivrii and Sigal finally managed to apply Fourier
integral operator methods to the atomic problem and proved the Scott
correction for molecules, which was recently extended to matter by Balodis 
Matesanz~\cite{Matesanz}.  

In \cite{Fefferman-Seco}, Fefferman and Seco gave a rigorous derivation
of the next correction
(after the Scott correction) in the asymptotics of the energy of atoms. 
This next correction had been predicted by Dirac~\cite{Dirac} 
and Schwinger~\cite{Schwinger}.  

As we shall explain below (see Page \pageref{page:classicalcoherent})
one cannot expect to be able to derive the Scott correction using the 
traditional method of coherent states. 

In this paper we introduce a new semiclassical approach 
generalizing the method of coherent states and show that this
approach can be used to give a fairly simple derivation of
the Scott correction for molecules. 

The standard coherent states method is based on representing operators
on $L^2(\R^n)$ as integrals of the form
\begin{equation}\label{eq:classicalcoherentrepresentation}
  \int_{\R^{2n}} a(u,q)\Pi_{u,q}\frac{dudq}{(2\pi h)^{n}},
\end{equation}
where $a(u,q)$ is a function (the symbol of the operator) on the
classical phase space $\R^{2n}$ and $\Pi_{u,q}$ is a non-negative operator with
the properties  
$$
 \Tr\Pi_{u,q}=1, \quad \int_{\R^{2n}} \Pi_{u,q}\frac{dudq}{(2\pi h)^{n}}={\bf 1}.
$$

For the classical coherent states $\Pi_{u,q}$ is the one-dimensional
projection $\left|u,q\right\rangle\left\langle u,q\right|$ 
onto the normalized function 
\begin{equation}\label{eq:classicalcoherentstates} 
  \langle x|u,q\rangle = (\pi h)^{-n/4} e^{-(x-u)^2/2h} e^{iqx/h}.
\end{equation}

We generalize this by representing operators in the form 
\begin{equation}\label{form}
   \int {\cal G}_{u,q}\,{\widehat A}_{u,q}\,{\cal G}_{u,q}\frac{dudq}{(2\pi h)^{n}}. 
\end{equation}
Here ${\cal G}_{u,q}$ is some self-adjoint operator such that 
its square plays the role of $\Pi_{u,q}$ and
${\widehat A}_{u,q}=B_0(u,q)+B_1(u,q)\cdot {\x}-ih B_2(u,q)\cdot\nabla$ 
is a differential operator linear in ${\x}$ and $-ih\nabla$. (We have
denoted by $\x$ the position operator.) We shall
make an explicit choice of $\cG_{u,q}$ in Sect.~\ref{sec:coherent}.
In other words, we allow the symbol in the coherent state operator 
representation to be not just a real function on phase space but to
take values in first order differential operators. If we consider, for example,
a Schr\"odinger operator of the form $-h^2\Delta+V(\hat{x})$, where a natural choice 
of the coherent state symbol would be $a(u,q)=q^2+V(u)$, then 
the new idea is now  to choose the linear approximation
$$
 {\widehat A}_{u,q}=a(u,q)+\partial_u a(u,q)(\x-u)+\partial_q a(u,q)(-ih\nabla-q).
$$
The representation (\ref{form}) will then be a better approximation of
the Schr\"odinger operator than
(\ref{eq:classicalcoherentrepresentation}) (see
Theorem~\ref{thm:coherentrepresentation} for details).

In order to explain the Scott correction
we consider the non-relativistic Schr\"odinger operator for 
a neutral molecule
\[ H(Z,R)=H(Z_1,\ldots,Z_M;R_1,\ldots, R_M)
    =\sum_{i=1}^Z\left(-\mfr{1}{2}\D_i - V(Z,R,x_i)\right)
     +\sum_{i<j}\frac{1}{|x_i-x_j|}.
\]
We have $|Z|=\sum_{j=1}^M Z_j$ electrons of charges $e=-1$ and $M$ nuclei of charges
$Z_j$ located at the fixed positions $R_1,\ldots, R_N$; we use atomic units where 
$\hbar^2=m$. The number $M$ is an arbitrary but fixed integer throughout this paper. 
The potential
\be\label{V} 
   V(Z,R,x) = \sum_{j=1}^M \frac{Z_j}{|x-R_j|}
\ee
describes the interaction of a single electron with all the nuclei. 
The operator $H(Z,R)$ acts as an unbounded operator in the space 
$\bigwedge^N L^2(\R^3\times\{-1,1\})$, where $\pm1$ refer to the spin variables.
We are interested in the ground state energy 
$$ E(Z,R) = \inf\mbox{spec} H(Z,R)
$$
and, in particular, in the asymptotic expansion for large charges. 
In defining the energy we have ignored the nuclear repulsion. It would
simply shift the energy by a 
constant depending on $Z$ and $R$.

The version of the Scott correction that we prove in this paper can
now be stated as follows. 

\begin{thm}[Scott correction] \label{main theorem} 
Let $Z=|Z|(z_1,\ldots,z_M)$,
where $z_1,\ldots,z_M>0$ and $R=|Z|^{-1/3}(r_1,\ldots,r_M)$, where
$|r_i- r_j|>r_0$ for some $r_0>0$. Then,
\begin{equation} 
  E(Z,R) = E^{\rm TF} (Z,R) + \mfr{1}{2} \sum_{1\le j\le M} Z_j^2 + 
  {\cal O} (|Z|^{2-1/30}),
\end{equation}
as $|Z|\to\infty$, where the error term ${\cal O} (|Z|^{2-1/30})$
besides $|Z|$ depends only on $z_1,\ldots,z_M$, and $r_0$. 
\end{thm}
This is established in lemmas \ref{lower bound} and \ref{upper bound}.
In fact, one could improve slightly on the error estimate to the
expense of limiting the range of $Z$ and $R$, and vice verse. 

It turns out that $E^{\rm TF} (Z,R) $ is of order $|Z|^{7/3}$ and
the next term $\mfr{1}{2} \sum_{1\le j\le M} Z_j^2$ is the Scott
correction. 

Part of our derivation of Theorem~\ref{main theorem} is similar to the
multi-scale analysis in \cite{Ivrii-Sigal} and we adopt their notation.
Our semiclassical method, however, is very different. It does not rely on the
spectral calculus, but uses only the quadratic form representation of
operators. 
Moreover, we treat the Coulomb singularities completely differently
from \cite{Ivrii-Sigal}. In treating the singularities and the region
near infinity the Lieb-Thirring inequality plays an essential role. 

Another virtue of our proof is that it gives an explicit trial state
for the energy that is correct to an order including the Scott
correction. This is, in fact, how we prove that the Scott correction
is correct as an asymptotic upper bound.

This paper is organized as follows. In Sect.~\ref{sec:tools} 
we list for the convenience of the reader the analytic tools that 
we shall use in a crucial way. In Sect.~\ref{sec:tf} we review Thomas-Fermi theory.
In Sect.~\ref{sec:coherent} we introduce the new coherent states.
In Sect.~\ref{sec:sc} we apply this new tool to 
prove the semi-classical expansion of the sum of the negative eigenvalues of a non-singular 
Schr\"odinger operator localized in some bounded region of space. This
is the key application of our new method.
The proof for the semi-classical expansion for the Thomas-Fermi potential is
presented in Sect.~\ref{sec:sctf}. In Sect.~\ref{sec:scott}
we finally prove lower and upper bound for the 
molecular quantum ground state energy. Some calculations
concerning the new coherent states and a theorem on constructing a
particular partition of
unity are put into the appendices.  

\section{Preliminaries}

\subsection{Analytic tools}\label{sec:tools}

In this subsection we collect the main analytic tools  which we shall
use throughout  the paper. We 
do not prove them here but give the standard references. Various constants are typically 
denoted by the same letter $C$, although their value might, for instance, change from 
one to the next line.

Let $p\ge1$, then a complex-valued function $f$ (and only those will be considered here)
is said to be in $L^p(\R^n)$ if the norm 
$\|f\|_p := \left(\int |f(x)|^p \,dx\right)^{1/p}$ is finite. For any $1\le p\le t\le
q\le\infty$ we have the inclusion $L^p\cap L^q\subset L^t$, since by H\"older's inequality
$\| f\|_t \le \|f\|_p^\lambda \|f\|_q^{1-\lambda}$ with $\lambda p^{-1}+(1-\lambda) q^{-1}
=t^{-1}$.

We call $\c$ a density matrix on $L^2(\R^n)$ if it is a trace class
operator on $L^2(\R^n)$ satisfying the operator inequality ${\bf
  0}\le\c\le {\bf 1}$. The density of a density matrix $\gamma$ is the
$L^1$ function $\rho_\gamma$ such that
$\Tr(\gamma\theta)=\int\rho_\gamma(x)\theta(x)dx$ for all $\theta\in
C_0^\infty(\R^n)$ considered as multiplication operators.  

We also need an extension to many-particle states. Let $\psi\in
\bigotimes^N L^2(\R^3\times\{-1,1\})$ be an $N$-body wave-function.
Its one-particle density $\rho_\psi$ is defined by 
$$
\rho_\psi(x) = \sum_{i=1}^N \sum_{s_1=\pm1}\cdots\sum_{S_N=\pm1}\int
|\psi(x_1,s_1\ldots,x_N,s_N)|^2\,\delta(x_i-x)\,dx_1\cdots x_N.
$$
The next inequality we recall is crucial to most of our estimates.

\begin{thm}[Lieb-Thirring inequality] \label{Lieb-Thirring}
 
{\bf One-body case:} Let $\c$ be a density operator on $L^2(\R^n)$, 
then we have the Lieb-Thirring inequality 
\begin{equation}\label{LTdensity}
   \Tr\left[-\mfr{1}{2}\Delta \gamma\right]\geq K_n\int\rho_\gamma^{1+2/n}
\end{equation}
with some positive constant $K_n$.  Equivalently, let $V\in L^{1+n/2}(\mathbb R^n)$ 
and $\c$ a density operator, then \be\label{LT} \Tr
[(-\mfr{1}{2}\D + V)\c] \ge -L_n \int |V_-|^{1+n/2}, \ee where $x_- :=
\min\{x,0\}$, and $L_n$ some positive constant.

{\bf Many-body case:} Let $\psi\in \bigwedge^N
L^2(\R^{3}\times\{-1,1\})$. Then,
\begin{equation}\label{eq:LTmbcase}
  \left\langle\psi,\sum_{i=1}^N-\mfr{1}{2}\Delta_i\psi\right\rangle\geq 
  2^{-2/3} K_3\int\rho_\psi^{5/3} .
\end{equation}
\end{thm}

The original proofs of these inequalities can be found in
\cite{Lieb-Thirring}. 

From the min-max principle it is clear
that the right side of (\ref{LT}) is in fact a lower bound on the sum
of the negative eigenvalues of the operator $-\frac{1}{2}\Delta + V$.

We shall use the following standard notation:
$$D(f)=D(f,f)=\frac{1}{2}\int\!\!\int \bar{f}(x)|x-y|^{-1} f(y)\, dx dy .
$$
It is not difficult to see (by Fourier transformation) that $\|f\| := D(f)^{1/2}$ is a norm.

\begin{thm}[Hardy-Littlewood-Sobolev inequality] There exists a constant $C$ such that
\be \label{Hardy-Littlewood-Sobolev} 
    D(f)\le C\,\| f \|_{6/5}^2.
\ee
\end{thm}
The sharp constant $C$ has been found by Lieb \cite{Lieb:sob}, see also \cite{Lieb-Loss}. 

In order to localize into different regions of space we shall use the standard
IMS-formula
\begin{equation}\label{eq:IMS}
  -\mfr{1}{2}\theta^2\Delta-\mfr{1}{2}\Delta\theta^2=-\theta\Delta\theta-(\nabla\theta)^2 ,
\end{equation}
which holds, by a straightforward calculation, for all bounded $C^1$-functions $\theta$
(here considered as a multiplication operator).

Finally we state the two inequalities which we need to estimate the many-body ground state
energy $E(Z,R)$ by an energy of an effective one-particle quantum system. 
The first one is an electrostatic inequality providing us with a lower bound. 
This inequality is due to Lieb \cite{Lieb3}, and was improved in \cite{Lieb-Oxford}. 

\begin{thm}[Lieb-Oxford inequality] Let $\psi\in L^2(\R^{3N})$ be normalized, and $\rho_\psi$ its 
one-electron density. Then,
\be\label{Lieb-Oxford} \left\langle \psi,\sum_{1\le i<j\le N} |x_i -x_j|^{-1} 
   \psi\right\rangle \ge D(\rho_\psi) - C \int \rho_\psi^{4/3} .
\ee
\end{thm}
The best-known numerical value is 1.68, but this does not play a role
here. 

An upper bound to $E(Z,R)$ is provided by a variational principle for 
Fermionic systems. This is also due to Lieb \cite{Lieb2}. 

\begin{thm}[Lieb's Variational Principle] \label{Lieb's Variational Principle}
Let $\c$ be a density matrix on $L^2(\R^3)$ satisfying
$2\Tr\c= 2\int \rho_\c(x)\, dx \leq Z$ (i.e., less than or equal to the number of
electrons) with the kernel $\rho_\c(x)=\c(x,x)$. Then 
\be
\label{variational principle} E(Z,R)\le 2\Tr\Big[\left(-\mfr{1}{2}\D -
  V(Z,R,x)\right)\c\Big] + D(2\rho_\gamma) .  
\ee 
\end{thm}
The factors 2 above are due to the spin degeneracy.

\subsection{Thomas-Fermi theory }\label{sec:tf}

Consider ${\mathbf z}=(z_1,\ldots,z_M)\in\R_+^M$ and 
${\mathbf r}=(r_1,\ldots,r_M)\in\R^{3M}$.  Let
$0\le\rho\in L^{5/3}(\mathbb R^3)\cap L^1(\mathbb R^3)$ then the
Thomas-Fermi (TF) energy functional, ${\cal E}^{{\rm TF}}$, is
defined as 
$$ 
{\cal E}^{{\rm TF}}(\rho) =
\mfr{3}{10}(3\pi^2)^{2/3}\int \rho(x)^{5/3}\, dx - \int
V({\mathbf z},{\mathbf r},x)\rho(x)\, dx+ D(\rho) ,
$$
where $V$ is as in (\ref{V}).

By the Hardy-Littlewood-Sobolev inequality the Coulomb energy,
$D(\rho)$, is finite for functions $\rho\in L^{5/3}(\mathbb R^3)\cap
L^1(\mathbb R^3)\subset L^{6/5}(\mathbb R^3)$. Therefore, the TF-energy
functional is well-defined. Here we only state the properties about
TF-theory which we use throughout the paper without proving them. The
original proofs can be found in \cite{Lieb-Simon} and \cite{Lieb1}.

\begin{thm}[Thomas-Fermi minimizer] 
For all ${\mathbf z}=(z_1,\ldots,z_M)\in \R_+^M$ and 
${\mathbf r}=(r_1,\ldots,r_M)\in\R^{3M}$ there exists a unique non-negative 
$\rho^{{\rm TF}}({\mathbf z},{\mathbf r},x)$ 
such that $\int \rho^{{\rm TF}}({\mathbf z},{\mathbf r},x)\,
dx=\sum_{k=1}^Mz_k$ and
$$
{\cal E}^{{\rm TF}}(\rho^{{\rm TF}}) = \inf
\left\{{\cal E}^{{\rm TF}} (\rho)\ :\ 0\le\rho\in L^{5/3}(\R^3)\cap
  L^1(\R^3)\right\} .
$$
We shall denote by 
$E^{{\rm TF}}({\mathbf z},{\mathbf r}) := {\cal E}^{{\rm TF}} (\rho^{{\rm TF}})$ 
the TF-energy. Moreover, let 
\begin{equation}\label{eq:tfpotgeneral}
 V^{{\rm TF}}({\mathbf z},{\mathbf r},x) := V({\mathbf
    z},{\mathbf r},x)
  -\rho^{{\rm TF}} * |x|^{-1}.
\end{equation}
be the TF-potential, then $V^{{\rm TF}}>0$ and $\rho^{{\rm TF}}>0$, 
and $\rho^{{\rm TF}}$ is the unique solution in 
$L^{5/3}(\R^3)\cap L^1(\R^3)$ to the TF-equation: 
\begin{equation}\label{eq:tfeqgeneral} 
   V^{{\rm TF}}({\mathbf z},{\mathbf r},x) =
   \mfr{1}{2}(3\pi^2)^{2/3}
   \rho^{{\rm TF}}({\mathbf z},{\mathbf r},x)^{2/3}.
\end{equation}
\end{thm}

Very crucial for a semi-classical approach is the {\it scaling}
behavior of the TF-potential.  It says that for any positive parameter
$h$ 

\begin{eqnarray}\label{scaling} 
  V^{{\rm TF}}({\mathbf z},{\mathbf r},x) &=&
  h^{-4} V^{{\rm TF}}(h^{3}{\mathbf z},h^{-1}{\mathbf r},h^{-1}x),
  \\
  \label{scaling:rho}
  \rho^{{\rm TF}}({\mathbf z},{\mathbf r},x) &=& h^{-6}\rho^{{\rm
      TF}}(h^{3}{\mathbf z},h^{-1}{\mathbf r},h^{-1}x)
  \\ 
  E^{\rm TF}({\mathbf z},{\mathbf r})&=&
  h^{-7}E^{\rm TF}(h^{3}{\mathbf z},h^{-1}{\mathbf r}).  
\end{eqnarray}
By $h^{-1}{\mathbf r}$ we mean that each coordinate is scaled by
$h^{-1}$, and likewise for $h^{3}{\mathbf z}$. 
By the TF-equation (\ref{eq:tfeqgeneral}), the
equations (\ref{scaling}) and (\ref{scaling:rho}) are obviously equivalent.  
Notice that the Coulomb-potential, $V$, has the claimed scaling behavior. 
The rest follows from the uniqueness of the solution of the TF-energy
functional.

We shall now establish the crucial estimates that we need about the TF
potential. Let 
\begin{equation}\label{ddefinition}
  d(x)=\min\{|x-r_k|\ |\ k=1,\ldots,M\}
\end{equation}
and
\begin{equation}\label{fdefinition}
  f(x)=\min\{d(x)^{-1/2}, d(x)^{-2}\}.
\end{equation}
For each $k=1,\ldots,M$ we define the function
\begin{equation}\label{eq:Wdefinition}
  W_k({\mathbf z},{\mathbf r},x)=V^{\rm TF}({\mathbf z},{\mathbf r},x)
  -z_k|x-r_k|^{-1}.
\end{equation}
The function $W_k$ can be continuously extended to $x=r_k$. 
 
The first estimate in the next theorem is very similar to
a corresponding estimate in \cite{Ivrii-Sigal}.
\begin{thm}[Estimate on $V^{\rm TF}$]\label{thm:tfestimate}
Let ${\mathbf z}=(z_1,\ldots,z_M)\in \R_+^M$ and
${\mathbf r}=(r_1,\ldots,r_M)\in \R^{3M}$.
For all multi-indices $\alpha$ and all $x$ with 
$d(x)\ne0$ we have  
\begin{equation}\label{eq:tfdf}
\left|\partial^\alpha_x\V({\mathbf z},{\mathbf r},x)\right|\leq C_\alpha f(x)^2
d(x)^{-|\alpha|}, 
\end{equation}
where $C_\alpha>0$ is a constant which depends on $\alpha$, $z_1,\ldots,z_M$,
and $M$.

Moreover, for $|x-r_k|<r_{\min}/2$, where 
$r_{\min}=\min_{k\ne\ell}|r_k-r_\ell|$ we have 
\begin{equation}\label{eq:westimate}
 0\leq W_k({\mathbf z},{\mathbf r},x)
  \leq Cr_{\min}^{-1}+C,
\end{equation}
where the constants $C>0$ here depend on $z_1,\ldots,z_M$,
and $M$. 
\end{thm}
\begin{proof}
Throughout the proof we shall denote  all constants that
depend on $\alpha$, $z_1,\ldots,z_M$, $M$ by $C_\alpha$. 
Constants that depend on $z_1,\ldots,z_M$ we denote by $C$.
In this proof we shall omit the dependence on ${\mathbf r}$ and
${\mathbf z}$ and simply write $\V(x)$ and $W_k(x)$.

We proceed by induction over $|\alpha|$. If $\alpha=0$ we have the well 
known bound \cite{Lieb-Simon}
that 
\begin{equation}\label{eq:LSmolecule}
 0\leq \max\{\V_{r_k}(x)\ |\ k=1,\ldots,M\}\leq \V(x)\leq \sum_{k=1}^M\V_{r_k}(x),
\end{equation}
where $\V_{r_k}$ denotes the Thomas-Fermi potential of a neutral
atom with a nucleus placed at $r_k\in\R^3$ with nuclear charge $z_k$.
This potential satisfies the bounds \cite{Lieb-Simon}
\begin{equation}\label{eq:LSatom}
  C_-\min\{z_k|x-r_k|^{-1}, |x-r_k|^{-4}\} \leq \V_{r_k}\leq 
  C_+\min\{z_k|x-r_k|^{-1}, |x-r_k|^{-4}\},             
\end{equation}
where $C_\pm>0$ are universal constants (note that by scaling
(\ref{scaling})
it is enough to consider the case $z_k=1$). We therefore get that
\begin{equation}\label{eq:tfphi}
 C_-\min\{z_1,\ldots,z_M,1\} f(x)^2\leq \V(x)
 \leq C_+ M \max\{z_1,\ldots,z_M,1\} f(x)^2.
\end{equation}
This in particular gives  (\ref{eq:tfdf})  for $\alpha=0$.
Assume now that (\ref{eq:tfdf}) has been 
proved for all multi-indices
$\alpha$ with $|\alpha|<M$, for some $M>0$. 
We shall first establish an estimate
for the derivatives $\partial^{\alpha}\rho$ of the TF density $\rho$.

{F}rom the TF equation we have that 
$\rho=C(\V)^{3/2}$. Thus $\partial^{\alpha}\rho(x)$ is a sum of terms of
the form 
$$
\V(x)^{3/2-k}\partial^{\beta_1}\V(x)\cdots \partial^{\beta_k}\V(x)
$$
where $k=0,\ldots,|\alpha|$ and $|\beta_1|+\ldots+|\beta_k|=|\alpha|$.
Thus by the induction hypothesis and (\ref{eq:tfphi}) we have for
$|\alpha|< M$ that 
\begin{equation}\label{eq:tfrho}
\left|\partial^{\alpha}\rho(x)\right|\leq
C_\alpha
f(x)^3d(x)^{-|\alpha|}.
\end{equation}

We now turn to the potential. Given $\alpha$ with $|\alpha|=M$.
Choose some decomposition $\alpha=\beta+\alpha'$, where $|\beta|=1$
and $|\alpha'|=M-1$.

For all $y$ such that
$|y-x|<d(x)/2$ we write 
$$
\partial^{\alpha'}\V(y)=-\int_{|u-x|<d(x)/2}
\partial^{\alpha'}\rho(u)|y-u|^{-1}du+R(y),
$$
where 
$$
R(y)=\partial^{\alpha'}\V(y)+\int_{|u-x|<d(x)/2}
\partial^{\alpha'}\rho(u)|y-u|^{-1}du
$$
is a harmonic function of $y$ for $|y-x|<d(x)/2$, since
$\Delta\V(y)=4\pi\rho(y)$ for such $y$. It follows that
$$
\left|\partial^{\beta}R(x)\right|\leq C
d(x)^{-1}\sup_{|\xi-x|=d(x)/2}|R(\xi)|.
$$
This can be seen by the Poisson formula
$$
R(y)= \int\limits_{|\xi-x|=d(x)/2} 
\frac{\mfr{d(x)^2}{4}-|y-x|^2}{2\pi d(x)|\xi-y|^{3}}R(\xi)\,dS(\xi),
$$
valid for $|y-x|<d(x)/2$. Here $dS$ denotes the surface measure of
$\{\xi\,:\, |\xi-x|=d(x)/2\}$.
To estimate $\sup_{|\xi-x|=d(x)/2}|R(\xi)|$ we use the induction
hypothesis, i.e., (\ref{eq:tfdf})
and (\ref{eq:tfrho}) for $\alpha'$. Note that 
for $|\xi-x|\leq d(x)/2$ we have that 
$d(x)/2\leq d(\xi)\leq 3d(x)/2$ and hence also that 
$f(\xi)\le 4 f(x)$.
Thus, if we also use that $f(x)\leq d(x)^{-2}$  we get that  
$$
 \sup_{|\xi-x|=d(x)/2}|R(\xi)|\leq C_{\alpha'} d(x)^{-|\alpha'|}f(x)^{2}
$$
and hence that 
$$
 \left|\partial^{\beta}R(x)\right|\leq C_\alpha d(x)^{-|\alpha|}f(x)^2.
$$
Finally we have that 
$$
\left|\partial^{\beta}_y\int_{|u-x|<d(x)/2}
  \partial^{\alpha'}\rho(u)|y-u|^{-1}du\right|
\leq
 \int_{|u-x|<d(x)/2}
 \left|\partial^{\alpha'}\rho(u)\right||y-u|^{-2}du
$$
The estimate (\ref{eq:tfdf}) follows since 
$$\int_{|u-x|<d(x)/2}
 \left|\partial^{\alpha'}\rho(u)\right||x-u|^{-2}du \leq
 C_{\alpha'} f(x)^3d(x)^{-|\alpha'|+1}\leq C_\alpha f(x)^2d(x)^{-|\alpha|},
$$
where we have again used that $f(x)\leq d(x)^{-2}$. 

The estimate (\ref{eq:westimate}) follows from (\ref{eq:LSmolecule})
and (\ref{eq:LSatom}) if we note that the atomic potential satisfies 
$$
 0\leq V^{\rm TF}_{r_k}(x)-z_k|x-r_k|^{-1}=\rho^{\rm TF}_{r_k}*|x|^{-1}\leq C.
$$ 
Here $\rho^{\rm TF}_{r_k}$ is the atomic density.
These last estimates follow since $\rho^{\rm TF}_{r_k}$ is non-negative and bounded in
$L^{5/3}$ and in $L^1$. 
\end{proof}

The relation of Thomas-Fermi theory to semiclassical analysis is that
the semiclassical density of a gas of non-interacting electrons moving 
in the Thomas-Fermi potential $\V$ is simply the Thomas-Fermi density.
More precisely, the semiclassical approximation to 
the density of the projection onto the eigenspace corresponding to 
negative eigenvalues of the Hamiltonian $-\frac{1}{2}\Delta-\V$ is   
$$
2\int_{\frac{1}{2}p^2-\V({\mathbf z},{\mathbf r},x)\le0}1
\frac{dp}{(2\pi)^3} = 2^{3/2} (3\pi^2)^{-1} (\V)^{3/2}({\mathbf
  z},{\mathbf r},x)= \rho^{\rm TF}({\mathbf z},{\mathbf r},x).
$$ 
Here the factor two on the very left is due to the spin degeneracy.
Similarly, the semiclassical approximation to the energy of the gas,
i.e., to the sum of the negative eigenvalues of
$-\frac{1}{2}\Delta-\V$ is  
\begin{equation}\label{eq:sc=tf}
  2\int \left(\frac{1}{2}p^2-\V({\mathbf z},{\mathbf r},x)\right)_-
  \frac{dpdx}{(2\pi)^3}= -{\frac {4\sqrt {2}}{15{\pi }^{2}}}\int
  \,\V({\mathbf z},{\mathbf r},x)^{5/2}dx=E^{{\rm TF}}({\mathbf
    z},{\mathbf r})+D(\rho^{\rm TF}).
\end{equation}
In Section~\ref{sec:sctf} we shall make the semiclassical
approximation more precise.

\section{The new coherent states}\label{sec:coherent}

We shall now define the new coherent states (or better coherent
operators\footnote{This should not be confused with the quantum coherent operators
introduced by Lieb and Solovej in \cite{Lieb-Solovej} in order to compare two quantum
systems.}) discussed in the introduction. I.e., we shall define the
operator $\cG_{u,q}$ that is used to represent operators in the form
(\ref{form}). 

The classical coherent states (\ref{eq:classicalcoherentstates})
localize in both $x$ and $p$ on a scale of order $h$. 
We shall choose $\cG_{u,q}$ to localize on a longer scale. We define
\begin{equation} \label{new coherent states} 
     {\cal G}_{u,q}:=\left(\frac{a}{\pi(1-ha)}\right)^n \int e^{-a/(1-ha)\, 
     [(u-u')^2+(q-q')^2]} \, |u',q'\rangle\langle u',q'|\, du' dq' .
\end{equation}
The new scale is $1/a>h$. 
Note that if we let $a\to 1/h$ then ${\cal G}_{u,q}^2$ 
converges to the projection $\Pi_{u,q}=|u,q\rangle \langle u,q|$.
A straightforward calculation gives the following result.

\begin{lemma}[Completeness of new coherent states]
These new coherent operators satisfy 
$$
  \int {\cal G}_{u,q}^2 \,\frac{dq}{(2\pi h)^n} =G_b(\x-u),\quad
  \int {\cal G}_{u,q}^2 \,\frac{du}{(2\pi h)^n} =G_b(-ih\nabla-q),
$$
where $\x$ denotes the operator multiplication by the position
variable $x$. Here $G_b(v)=(b/\pi)^{n/2}\exp(-bv^2)$ with
$b=2a/(1+h^2a^2)$. Note that $G_b$ has integral 1 and hence
$$
  \int {\cal G}_{u,q}^2 \,\frac{dudq}{(2\pi h)^n} = {\bf 1} .
$$
\end{lemma}

We shall study operators that can be written in the form (\ref{form}).
If ${\widehat A}_{u,q}=B_0(u,q)+B_1(u,q)\cdot {\x}-ih
B_2(u,q)\cdot\nabla$ 
is the operator valued symbol in (\ref{form}) we shall denote by 
$A_{u,q}$ the linear function 
$A_{u,q}(v,p)=B_0(u,q)+B_1(u,q)\cdot v+B_2(u,q)\cdot p$. 
When $A_{u,q}(v,p)$ is independent of $(v,p)$, i.e., if $B_1=B_2=0$ 
and if $a\to h^{-1}$ we recover the
usual coherent states representation of an operator. Thus, on the one hand
we do not use as sharp a phase space localization as the
one-dimensional coherent state projection since $a<1/h$, but  
on the other hand, we use a better approximation than if $A_{u,q}$ were just a constant. 

More generally we shall consider operators of the form 
\begin{equation}\label{eq:formf}
  \int {\cal G}_{u,q}\,f({\widehat A}_{u,q})\,{\cal G}_{u,q}dudq, 
\end{equation}
 where $f:\R\to\R$ is any polynomially bounded real function. As we
shall see in the next theorem the integrand above is a traceclass
operator for each $(u,q)$. The integral above is to be understood in
the weak sense, i.e., as a quadratic form. We shall consider
situations where the integral defines bounded or unbounded operators. 
\begin{thm}[Trace identity]\label{thm:traceidentity} 
Let $f:\R\to\R$ and $V:\R^n\to\R$  be polynomially bounded,
real measurable functions and 
$$\hat{A}=B_0 + B_1{\x} -ih B_2\nabla$$ 
a first order self-adjoint differential operator\footnote{The
  operator $\hat{A}$ is essentially self-adjoint on Schwartz functions on $\R^n$.}
with $B_0\in\R,B_{1,2}\in\R^n$. Then  
${\cal G}_{u,q}\,f(\hat{A}) \,{\cal G}_{u,q}\,V({\x})$ is a
trace class operator (when extended from $C_0^\infty(\R^n)$) and
  \begin{eqnarray*}
    {\Tr}\big[{\cal G}_{u,q}\,f(\hat{A}) \,{\cal G}_{u,q}\,V({\x})\big]
    &=& \int f(B_0+B_1v+B_2p)\,G_b(v-u)G_b(q-p)
    \\&&\,\times\,G_{(bh^2)^{-1}}(z)V(v+h^2ab(u-v)+z)dvdpdz.
  \end{eqnarray*}
In particular, $\Tr\Big[{\cal G}_{u,q}^2\Big]=1$.
\end{thm}
The proof is given in Appendix A. We shall need the following
extension of this theorem, where we however only give an estimate on
the trace. The proof is again deferred to Appendix A.
\begin{thm}[Trace estimates] \label{trace formula} Let $f,\hat{A}$ be
as in the previous theorem. Let moreover $\phi\in C^{n+4}(\mathbb R^n)$
be a bounded, real function with all derivatives up to order $n+4$ bounded
and $V,F\in C^2(\R^n)$ be real functions with bounded second derivatives.
Then, for $h<1$,
$1<a< 1/h$ and $b=2a/(1+h^2a^2)$ we have with 
$\s(u,q)=F(q)+V(u)$
that\footnote{The operator 
${\cal G}_{u,q}\,f(\hat{A}) \,{\cal G}_{u,q}\,
\phi(\hat{x}) \left(F(-ih\nabla) + V({\x})\right)\phi(\hat{x}) $ 
is originally defined on, say $C_0^\infty(\R^n)$, but it is part of the
claim of the theorem that it extends to a traceclass operator on
all of $L^2(\R^n)$.}
\begin{eqnarray*} \lefteqn{{\Tr}\big[{\cal G}_{u,q}\,f(\hat{A}) 
    \,{\cal G}_{u,q}\,
    \phi(\x) \left(F(-ih\nabla) + V({\x})\right)\phi(\x) \big]
    }
  \\
  &=&
  \int f(B_0+B_1v+B_2p)\,G_b(v-u)G_b(q-p)
  \\
  &&\times\,\Bigl[\Big(\phi(v+h^2ab(u-v))^2 + E_1(u,v)\Big)
  \s(v+h^2ab(u-v),p+h^2ab(q-p))\\&&{}
  \phantom{\times\,\Bigl[}+ E_2(u,v;q,p)\Bigr]\, dv dp ,
\end{eqnarray*}
with $\|E_1\|_\infty,\|E_2\|_\infty\le C h^2b$ where $C$ depends only on 
$$\sup_{|\nu|\le n+4} \|\p^\nu \phi\|_{\infty},\ 
  \sup_{|\nu|=2}\|\p^\nu
  V\|_\infty,\ \hbox{and}\ \sup_{|\nu|=2}\|\p^\nu
  F\|_\infty.
$$
(Note that the assumption $1<a<1/h$ implies $1<b<1/h$.) 
\end{thm} 
The above theorem shall be used to prove an upper bound on the sum of
eigenvalues of the operator $F(-ih\nabla) + V({\x})$, in the case when
$F(p)=p^2$. This is done in
Lemma~\ref{lm:upperbound} by constructing a trial density matrix in the 
form (\ref{eq:formf}). 

To prove a lower bound on the sum of the negative eigenvalues one 
approximates the Hamiltonian $F(-ih\nabla)+V(\x)$ 
by an operator represented in the form (\ref{form}).
This approximation which we now formulate is also proved in Appendix A.
\begin{thm}[Coherent states representation]\label{thm:coherentrepresentation}
Consider functions $F,V\in C^3(\R^n)$, for which all second and third 
derivatives are bounded. Let $\s(u,q)=F(q)+V(u)$, then we have for
$a<1/h$ and $b=2a/(1+h^2a^2)$ the 
representation
$$
F(-ih\nabla)+V(\x) = \int \,{\cal G}_{u,q}\widehat H_{u,q}
     {\cal G}_{u,q}\frac{du dq}{(2\pi h)^n} + {\mathbf E}
$$
(as quadratic forms on $C^\infty_0(\R^n)$), with the operator-valued symbol
\begin{equation}\label{eq:coherentrepresentation}
\widehat H_{u,q}= \s(u,q)+\frac{1}{4b}\D \s(u,q) 
+ \p_u \s(u,q)({\x}-u) + \p_q \s(u,q)(-ih\nabla -q).
\end{equation}
The error term, ${\mathbf E}$ is a bounded operator with
$$
\|\mathbf E\|\leq Cb^{-3/2}\sum_{|\alpha|=3}\|\partial^\alpha
\s\|_\infty+Ch^2b\sum_{|\alpha|=2}\|\partial^\alpha\s\|_\infty .
$$
\end{thm}

\section{Semi-classical estimate of
  Tr$\left[\phi(-h^2\Delta+V)\phi\right]_-$}\label{sec:sc}

In this Section we study the sum of the negative eigenvalues of
Schr\"odinger operators with regular potentials localized in a bounded
region by a localization function $\phi$.  This will turn out to be
the key theorem in this paper.  Let us recall our convention that
$x_-=(x)_- = \min\{x,0\}$. For convenience we consider balls, and we start
with the unit ball.

\begin{thm}[Local semiclassics] \label{local semi-classics}
  Let $n\geq3$ and $\phi\in C_0^{n+4}(\mathbb R^n)$, be supported in
  a ball $B\subset\R^n$ of
  radius $1$ and $V\in C^3(\overline{B})$ be a real function.  
  Let $H=-h^2\Delta+V$, $h>0$ and $\s(u,q)=q^2+V(u)$. Then,
  $$
  \left|{\Tr}[\phi H\phi]_-- (2\pi h)^{-n}\int \,\phi^2(u) \s(u,q)_- du dq\right|
  \leq Ch^{-n+6/5} .
  $$
  The constant $C>0$ here depends only on $n,\|\phi\|_{C^{n+4}}$ and
  $\|V\|_{C^3}$. [Here $\| V\|_{C^3}=
  \sup_{|\alpha|\leq3}\|\partial^\alpha V\|_\infty$.]
\end{thm}

With the classical coherent states the estimate one would normally prove
would be that the right side above is $C h^{-n+1}$. 
We find it instructive to sketch the proof of this here in order to 
make the comparison with the new method clearer. \label{page:classicalcoherent}

We may assume that $V$ is defined on all of $\R^n$ with bounded second 
and third order derivatives. We
shall  here assume that $h<1$.
{F}rom Theorem~\ref{thm:coherentrepresentation} with $a=1/h$ we have
$$H=-h^2\D +V = \int [q^2 +V(u)]\, |u,q\rangle\langle u,q|\, 
\frac{du dq}{(2\pi h)^n} + {\bf E} ,
$$
with $\|{\bf E}\| \le C h$. The constant depends on the second and
third order derivatives of $V$, which are bounded.  
We have here used that for $a=b=1/h$ the first order terms in $\x$ and 
$\nabla$ do not contribute (see (\ref{eq:firstorder}) below). Moreover, 
the term $\Delta\sigma$ is of order $h$ since $V$
has bounded second order derivatives. 
The error term ${\bf E}$ can be controlled using the Lieb-Thirring
inequality as in (\ref{eq:LTerror}) below.

Then from Theorem~\ref{trace formula} we have
\begin{eqnarray*} 
  {\Tr}[\phi H\phi]_- &\ge&\int[q^2+V(u)]_-\, 
  \Tr\Big[\phi|u,q\rangle\langle u,q|\phi\Big] 
  \frac{du dq}{(2\pi h)^{n}} -  Ch^{-n+1} 
  \\
  &=&\int[q^2+V(u)]_- \,\phi^2(u)\frac{du dq}{(2\pi h)^{n}} -  Ch^{-n+1} ,
\end{eqnarray*}      
where $C$ now also depends on the derivatives of $\phi$. 

For an upper bound we set
$$
\c = \int \chi_{(-\infty,0]}[q^2+V(u)]\,|u,q\rangle\langle u,q|\, \frac{du dq}{(2\pi h)^n},
$$
where $\chi_{(-\infty,0]}$ denotes the characteristic function of the
interval $(-\infty,0]$.  It is clear that ${\bf 0}\le\c\le{\bf 1}$.
When calculating $\Tr[\c \phi H\phi]$ we may again refer to the general
theorem \ref{trace formula} with $a=b=1/h$. We obtain
\begin{eqnarray*} 
  \Tr[\c \phi H\phi]&=&\int \chi_{(-\infty,0]}[q^2+V(u)] \,
  \Tr\Big[|u,q\rangle\langle u,q| \phi H\phi\Big]\, \frac{du dq}{(2\pi 
    h)^n}
  \\
  &\leq&\int [q^2+V(u)]_-\,\phi^2(u)\,\frac{du dq}{(2\pi h)^n} + C h^{-n+1}.  
\end{eqnarray*}

As mentioned in the Introduction
it is important that we obtain errors
bounded by $Ch^{-n+1+\varepsilon}$ for some $\varepsilon>0$ as in 
Theorem~\ref{local semi-classics}.
We shall prove Theorem~\ref{local semi-classics}  by again 
proving upper and lower bounds on ${\Tr}[\phi H\phi]_-$.

\begin{lemma}[Lower bound on ${\Tr}(\phi H\phi)_-$]
Let $n\geq3$, $\phi\in C_0^{n+4}(\mathbb R^n)$ be supported in a ball $B$ of radius
$1$ and assume that $V\in C^3(\overline{B})$.  
Let $H=-h^2\Delta+V$, $h>0$. Then,
$$ {\Tr}[\phi H\phi]_- 
  \ge (2\pi h)^{-n}\int \,\phi^2(u) \s(u,q)_- du dq
  -Ch^{-n+6/5} .
$$
The constant $C>0$ here depends only on $n,\|\phi\|_{C^{n+4}}$ and
$\|V\|_{C^3}$. 
\end{lemma}
\begin{proof}
Since $\phi$ has support in the ball $B$ we may 
without loss of generality assume that $V\in C^3_0(\R^3)$ with the
support in a ball $B_2$ of radius $2$
and that the norm $\| V\|_{C^3}$ refers to the supremum over all 
of $\R^n$. We shall not explicitly follow how the 
error terms depend on $\|\phi\|_{C^3}$ and $\|V\|_{C^3}$. 
All constants denoted by $C$ depend on $n,\|\phi\|_{C^3}$,
$\|V\|_{C^3}$.

First note that by the Lieb-Thirring inequality we have that 
$$ 
{\Tr}[\phi H\phi]_- \ge 
C\|\phi\|_\infty^2\int_{u\in B} \, \s(u,q)_-\frac{du dq}{(2\pi h)^n}
\geq-C h^{-n}.
$$
Consider some fixed $0<\tau<1$ (independent of $h$).
If $h\geq \tau$ then 
$$ {\Tr}[\phi H\phi]_- \ge \int \, \phi^2(u) \s(u,q)_-
\frac{du dq}{(2\pi h)^n}
   -C\tau^{-6/5}h^{-n+6/5}.
$$
We are therefore left with considering $h<\tau$. 
Of course one should really try to find the optimal value of $\tau$
(depending on $\phi$, and $V$) we shall however not do that. 
In studying the case $h<\tau$ it will be necessary to assume that 
the choice of $\tau$ is small enough. We 
therefore now assume that $h<\tau$ and that $\tau$ is small. 

{F}rom Theorem~\ref{thm:coherentrepresentation} we have that 
\begin{eqnarray}
  {\Tr}[\phi H\phi]_-&\geq& {\Tr}\left[\int \phi 
    \,{\cal G}_{u,q}\widehat{H}_{u,q}
    {\cal G}_{u,q}\phi\frac{du dq}{(2\pi h)^n}\right]_-\nonumber \\
  &&+ {\Tr}\left[\phi\left(-\varepsilon h^2\Delta
      -C(b^{-3/2}+h^2b)\right)
    \phi\right]_- \label{eq:LTerror}
\end{eqnarray}
where $0<\varepsilon<1/2$ and
$$
\widehat H_{u,q}=\widetilde{\s}(u,q)+\frac{1}{4b}\D \widetilde{\s}(u,q) 
+ \p_u\widetilde{\s}(u,q)({\x}-u) + \p_q \widetilde{\s}(u,q)(-ih\nabla -q)
$$
with $\widetilde{\s}(u,q)=(1-\varepsilon)q^2+V(u)$. We shall choose
$a$ depending on $h$ satisfying $\tau^{-1}\leq a<h^{-1}$ and hence
$\tau^{-1}\leq b<h^{-1}$.  It is clear (e.g. from the Lieb-Thirring
inequality) that the second trace above is estimated below by 
$-Ch^{-n} \varepsilon^{-n/2}(b^{-3/2}+h^2b)^{1+n/2} $. We shall choose
$\varepsilon=\frac{1}{4}(b^{-3/2}+h^2b)$; note that $\varepsilon<1/2$. 
Thus we find that the second trace is estimated by   
$-Ch^{-n}(b^{-3/2}+h^2b)$. From the variational
principle we have
$$
  {\Tr}[\phi H\phi]_-\geq \int {\Tr}\left[\phi 
    \,{\cal G}_{u,q}\left[\widehat H_{u,q}\right]_-
    {\cal G}_{u,q}\phi\right]\frac{du dq}{(2\pi h)^n}  
  - C h^{-n}(b^{-3/2}+h^2b).
$$

We first consider the integral over $u$ outside the ball $B_2$ of
radius 2, where $V=0$. Using Theorem~\ref{thm:traceidentity} (with $V$
replaced by $\phi^2$) and $\int \phi^2\le C$, we get that this part of the integral is
\begin{eqnarray*}
  &\displaystyle\int_{u\not\in
    B_2}&\left[(1-\varepsilon)q^2+\frac{n}{2b}(1-\varepsilon) + 2(1-\varepsilon)q\cdot(p-q)
  \right]_-
  G_b(p-q)G_b(u-v)\\
  &&\times\,G_{(bh^2)^{-1}}(z)\phi(v+h^2ab(u-v)+z)^2dvdpdz
  \frac{du dq}{(2\pi h)^n}\\
  &\geq&C\int(1-\varepsilon)[p^2-(p-q)^2]_-G_b(p-q)\frac{dp dq}{(2\pi
    h)^n}\geq-C b^{-(n+2)/2}h^{-n},
\end{eqnarray*}
which for all dimensions $n$ is bounded below by $-C b^{-3/2}h^{-n}$.
Actually it is not difficult to see that
we could have inserted a factor $e^{-Cb}$ on the right of
this estimate since $u\notin B_2$ and $\phi$ is supported in $B_1$, but
we do not need this here.
 
For the integral over $u\in B_2$ we use Theorem~\ref{trace formula} with
$F=0$ and $V=1$ to obtain 
\begin{eqnarray}
  {\Tr}[\phi H\phi]_-\geq
 &\displaystyle\int_{u\in B_2}&
  \left(\phi\left(v+h^2ab(u-v)\right)^2+ C h^2 b\right)
  G_b(u-v)G_b(q-p) \nonumber\\
  &&\times
  \left[H_{u,q}(v,p)\right]_-\frac{du dq}{(2\pi
      h)^n}dpdv
    -Ch^{-n}(b^{-3/2}+h^2b)\label{eq:phiHphi} ,
\end{eqnarray}
where
$$
H_{u,q}(v,p)=\widetilde{\s}(u,q)+\frac{1}{4b}
  \Delta\widetilde{\s}(u,q) 
    + \partial_u\widetilde{\s}(u,q)(v-u) + \partial_q
    \widetilde{\s}(u,q)(p-q).
$$
The rest of the proof is simply an estimate of the integral in
(\ref{eq:phiHphi}). 
Note that by Taylor's formula for $\widetilde\sigma$ we have
\begin{eqnarray}
  H_{u,q}(v,p)
    &\geq&
    \widetilde{\s}(v,p)+\widetilde{\xi}_v(u-v,q-p)-
C|u-v|(b^{-1}+|u-v|^2)\label{eq:atildeapp},
\end{eqnarray}
where
$$
\widetilde{\xi}_v(u,q)
=\frac{1}{4b}
  \Delta\widetilde{\s}(v,0) -(1-\varepsilon)q^2
  -\frac{1}{2}\sum_{ij}\partial_i\partial_j
  V(v)u_iu_j.
$$
We have here used that $\Delta\widetilde{\s}(v,p) $ is independent of
$p$ and that 
$\left|\Delta\widetilde{\s}(v,0)-\Delta\widetilde{\s}(u,0)\right|\leq C|u-v|$.
Since $\|V\|_{C^3}<\infty$ and thus, in particular,
$\widetilde{\s}(v,p)\geq (1-\varepsilon)p^2 - C$ we easily get that
\begin{equation}\label{eq:h-n}
  \int
  G_b(u-v)G_b(q-p)
  \left[H_{u,q}(v,p)\right]_-dpdqdv\geq -C
\end{equation} 
and hence from (\ref{eq:phiHphi}) that 
\begin{eqnarray*}
  {\Tr}[\phi H\phi]_-\geq
 &\displaystyle\int_{u\in B_2}&
  \phi\left(v+h^2ab(u-v)\right)^2
  G_b(u-v)G_b(q-p) \nonumber\\
  &&\times
  \left[H_{u,q}(v,p)\right]_-\frac{du dq}{(2\pi
      h)^n}dpdv-Ch^{-n}(b^{-3/2}+h^2b).
\end{eqnarray*}
Here we have of course used the fact that the 
$u$-integration is over a bounded region. {F}rom now on we may however 
ignore the restriction on the $u$-integration. Using (\ref{eq:atildeapp}) 
we find after the simple change of variables $u\to u+v$ and $q\to q+p$ that
\begin{eqnarray*}
  {\Tr}[\phi H\phi]_-&\geq&
  \int\phi\left(v+h^2abu\right)^2G_b(u)G_b(q)\\&&\times\, 
  \left[\widetilde{\s}(v,p)+\widetilde{\xi}_v(u,q)-
    C|u|(b^{-1}+|u|^2)\right]_-\frac{du dq}{(2\pi
    h)^n}dpdv\\&&-Ch^{-n}(b^{-3/2}+h^2b).
\end{eqnarray*}
We now perform the $p$-integration explicitly.
Recall that $\widetilde{\s}(v,p)=(1-\varepsilon)p^2+V(v)$ and
that $\int(p^2+s)_-dp=-\frac{2}{n+2}\omega_n|s_-|^{(n/2)+1}$, where 
$\omega_n$ is the volume of the unit ball in $\R^n$. 
We get
\begin{eqnarray}
  {\Tr}[\phi H\phi]_-&\geq&-\frac{2\omega_n}{n+2}
  (1-\varepsilon)^{-\frac{n}{2}}
  \int\phi\left(v+h^2abu\right)^2G_b(u)G_b(q)\nonumber\\&&
  \phantom{-\frac{2\omega_n}{n+2}(1-\varepsilon)}
  \times\,
  \biggl|\left[V(v)+\widetilde{\xi}_v(u,q)-
    C|u|(b^{-1}+|u|^2)\right]_-\biggr|^{\frac{n}{2}+1}\frac{du dq}{(2\pi
    h)^n}dv\nonumber\\&&-Ch^{-n}(b^{-3/2}+h^2b).\label{eq:lowerexpansion}
\end{eqnarray}
By expanding we find that 
\begin{eqnarray*}
  \lefteqn{\biggl|\left[V(v)+\widetilde{\xi}_v(u,q)-
    C|u|(b^{-1}+|u|^2)\right]_-\biggr|^{\frac{n}{2}+1}}&&\\
  &\leq& |V(v)_-|^{\frac{n}{2}+1}
  -\left(\frac{n}{2}+1\right)|V(v)_-|^{\frac{n}{2}}\widetilde{\xi}_v(u,q)\\
  &&
  +C\left(\left|\widetilde{\xi}_v(u,q)\right|+
    C|u|(b^{-1}+|u|^2\right)^2+C|u|(b^{-1}+|u|^2).
\end{eqnarray*}
We have here used that since $n\geq 3$, the function 
$\R\ni x\mapsto|x_-|^{\frac{n}{2}+1}$ is $C^2$.
Hence 
\begin{eqnarray*}
  {\Tr}[\phi H\phi]_-&\geq&-\frac{2\omega_n}{n+2}
  (1-\varepsilon)^{-\frac{n}{2}}
  \int\phi\left(v+h^2abu\right)^2G_b(u)G_b(q)\\&&
  \phantom{-\frac{2\omega_n}{n+2}(1-\varepsilon)}
  \times\,\left(|V(v)_-|^{\frac{n}{2}+1}
  -\left(\frac{n}{2}+1\right)|V(v)_-|^{\frac{n}{2}}\widetilde{\xi}_v(u,q)
  \right)
  \frac{du dq}{(2\pi
    h)^n}dv\\&&-Ch^{-n}(b^{-3/2}+h^2b).
\end{eqnarray*}
We now expand $\phi^2$
$$
\left|\phi\left(v+h^2abu\right)^2-
\phi(v)^2 - h^2abu\cdot\nabla\left(\phi^2\right)(v)\right|\leq
Ch^4a^2b^2|u|^2\leq Ch^2b^2|u|^2 ,
$$
and use the crucial identities
$$
 \int \widetilde{\xi}_v(u,q)G_b(u)G_b(q)dudq=0\quad
 \hbox{ and } \int uG_b(u)du=0.
$$
We thus arrive at 
\begin{eqnarray*}
  (2\pi h)^{n}{\Tr}[\phi H\phi]_-&\geq&-\frac{2\omega_n}{n+2}
  (1-\varepsilon)^{-\frac{n}{2}}
  \int\phi(v)^2|V(v)_-|^{\frac{n}{2}+1}dv-C(b^{-3/2}+h^2b)\\
  &=&(1-\varepsilon)^{-\frac{n}{2}}
  \int\phi(v)^2\s(v,p)_-dvdp-C(b^{-3/2}+h^2b).
\end{eqnarray*}
The lemma follows if we choose $a=\max\{h^{-4/5},\tau^{-1}\}$  
and $\varepsilon=\frac{1}{4}(b^{-3/2} + h^2b)$. Recall that
$a\le b\le 2a$. Thus $b^{-3/2}\le a^{-3/2}\le h^{6/5}$ and $h^2b\le 2h^2a\le2\tau^{-1/5}h^{6/5}$.

\end{proof}

In order to prove an upper bound on ${\Tr}(\phi H\phi)_-$ we shall
use that for any  density matrix $\gamma$ (i.e., a traceclass operator with  ${\bf 0}
\leq\gamma\leq{\bf 1}$) we have from the variational principle that 
${\Tr}(\phi H\phi)_-\leq {\Tr}(\phi H\phi\gamma)$. Hence the upper bound
needed to prove Theorem~\ref{local
  semi-classics} is a consequence of the following lemma. 
\begin{lemma}[Construction of trial density matrix]\label{lm:upperbound} 
Let $n\geq 3$, 
$\phi\in C_0^{n+4}(\mathbb R^n)$ be supported in a ball $B$ of radius 1, and 
$V\in C^3(\bar{B})$. Let $H=-h^2\D +V$, $h>0$ and $\s(u,q)=q^2+V(u)$. Then
there exists a density matrix $\gamma$ on $L^2(\R^n)$ such that
\begin{equation}
  {\Tr}[\phi H\phi\gamma]\le \int \, \phi^2(u) \s(u,q)_-\frac{du
    dq}{(2\pi h)^n} + Ch^{-n+6/5} \label{eq:lemmaupper}.
\end{equation}
Moreover, the density of $\gamma$ satisfies
\begin{equation}\label{eq:rhogammaprop1}
\left|\rho_\gamma(x)-(2\pi h)^{-n}\omega_n
  \left|V(x)_-\right|^{n/2}\right|\leq Ch^{-n+9/10},
\end{equation}
for (almost) all $x\in B$ and
\begin{equation}\label{eq:rhogammaprop2}
\left|\int\phi(x)^2\rho_\gamma(x)dx-(2\pi h)^{-n}\omega_n\int\phi(x)^2
  \left|V(x)_-\right|^{n/2}dx\right|\leq Ch^{-n+6/5},
\end{equation}
where $\omega_n$ is the volume of the unit ball in $\R^n$.
The constants $C>0$ in the above estimates depend only on
$n, \|\phi\|_{C^{n+4}}$, and $\|V\|_{C^3}$.
\end{lemma}
\begin{proof} As in the lower bound we choose some fixed $0<\tau<1$. 
We have for $h\ge \tau$  that for some $C>0$ 
$$\int \phi^2(u) \s(u,q)_-\,\frac{du dq}{(2\pi h)^n}
  + C\tau^{-6/5} h^{-n+6/5}\geq 0$${and}$$
\left|(2\pi h)^{-n}\omega_n
  V(x)_-\right|^{n/2}\leq C\tau^{-6/5}h^{-n+6/5},
$$  
If $h\geq\tau$ we may therefore choose $\gamma=0$. 
We may therefore now assume that 
$h<\tau$ and if necessary that $\tau$ is small enough depending
only on $\phi$, and $V$. Also as in the lower bound we may assume that 
$V\in C_0^3(\R^n)$ with support in the ball $B_{3/2}$ concentric with
$B$ and of radius $3/2$.

In analogy to the previous proof for the lower 
bound we now
for each $(u,q)$ define an operator $\hat{h}_{u,q}$ by
$$
\hat{h}_{u,q}=\left\{\begin{array}{cl}
\s(u,q)+ \frac{1}{4b}\Delta\s(u,q) 
+ \p_u\s(u,q)({\x}-u) +
\p_q {\s}(u,q)(-ih\nabla -q)&,\hbox{if } u\in B_2\\
0&,\hbox{if } u\not\in B_2
\end{array}\right..
$$
The corresponding function is
$$
{h}_{u,q}(v,p)=\left\{\begin{array}{cl}
\s(u,q) + \frac{1}{4b}\Delta\s(u,q) 
+ \p_u\s(u,q)(v-u) + \p_q {\s}(u,q)(p -q)&,\hbox{if } u\in B_2\\
0&,\hbox{if } u\not\in B_2
\end{array}\right. .
$$
Recall that $b=2a/(1+h^2a^2)$ (i.e., in particular $a\leq b\leq 2a$)
and as in the lower bound we shall choose $a=\max\{h^{-4/5},\tau^{-1}\}$

Similar to (\ref{eq:atildeapp}) we have for $u\in B_2$ that 
\begin{eqnarray}
  \left|h_{u,q}(v,p)-\s(v,p)-\xi_{v}(u-v,q-p)\right|
\leq C|u-v|(b^{-1}+|u-v|^2),\label{eq:happ}
\end{eqnarray}
where
$$
 \xi_{v}(u,q)=\frac{1}{4b}\Delta\s(v,0)-q^2
       -\mfr{1}{2}\sum_{i,j}\p_i\p_j
     V(v)u_iu_j.
$$
We have here used that $\Delta\s(v,p)$ is independent of $p$.

If we let $\chi=\chi_{(-\infty,0]}$ be the 
characteristic function of $(-\infty,0]$ we now define
\be \label{trial density}
   \c = \int {\cal G}_{u,q}\,\chi\big[\hat{h}_{u,q} \big]\,
             {\cal G}_{u,q}\,\frac{dudq}{(2\pi h)^n} .
\ee 
Since ${\bf 0}\le\chi\big[\hat{h}_{u,q}\big]\le\bf1$ it is obvious that ${\bf 0}
\le\c\le{\bf 1}$. Moreover, by Theorem~\ref{thm:traceidentity}
and (\ref{eq:happ}), $\gamma$ is easily seen to be a
traceclass operator with density
$$
  \rho_\gamma(x)=
  \int\chi\left(h_{u,q}(v,p)\right)G_b(u-v)G_b(p-q)G_{(bh^2)^{-1}}(x-v-h^2ab(u-v))dvdp
  \frac{dudq}{(2\pi h)^n}.
$$
If we change variables $u\to u+v$, $q\to q+p$ and
perform the $p$-integration we find that 
\begin{eqnarray}\label{eq:rhogamma}
  \rho_\gamma(x)&=&
  \omega_n\int_{u\in B_2-v} \Xi(v,u,q)G_b(u)G_b(q)
  G_{(bh^2)^{-1}}(x-v-h^2abu)dv
  \frac{dudq}{(2\pi h)^n}\nonumber\\
  &=&\omega_n\hspace{-20pt}\int\limits_{(1-h^2ab)u\in B_2-v}\hspace{-20pt}
  \Xi(v-h^2abu,u,q)G_b(u)
  G_b(q)G_{(bh^2)^{-1}}(x-v)dv\frac{dudq}{(2\pi h)^n}
\end{eqnarray}
where $\Xi(v,u,q)=\omega_n^{-1}\int\chi(h_{(u+v,q+p)}(v,p))\,dp\geq0$. From equation 
(\ref{eq:happ}) we have 
\begin{eqnarray}
  \biggl|\Xi(v,u,q)^{2/n}-\biggl|\biggl(V(v)+\xi_v(u,q)
  \biggr)_-\biggr|\biggr|\leq C|u|(b^{-1}+|u|^2),
  \label{eq:Westimate}
\end{eqnarray}
for all $v,q\in\R^n$ and $u\in B_2-v$. Since
$$
|\xi_v(u,q)-\xi_{v-h^2abu}(u,q)|\leq C h^2ab|u|(b^{-1}+|u|^2)
$$
we therefore also have
\begin{eqnarray*}
  \Biggl|\Xi(v-h^2abu,u,q)^{2/n}-\biggl|\biggl(V(v)+\eta_v(u,q)
  \biggr)_-\biggr|\Biggr|&\leq& Ch^4a^2b^2|u|^2\\
  &&{}+ C(1+h^2ab)|u|(b^{-1}+|u|^2),
\end{eqnarray*}
where 
$$
 \eta_v(u,q)=\xi_v(u,q)-h^2ab\nabla V(v)\cdot u .
$$
Hence from (\ref{eq:rhogamma})
\begin{eqnarray}
  \Biggl|\rho_\gamma(x)^{\frac{2}{n}}
  -\biggl(\omega_n\hspace{-20pt}\int\limits_{(1-h^2ab)u\in B_2-v}\hspace{-20pt}
  \biggl|\biggl(V(v)+\eta_v(u,q)
  \biggr)_-\biggr|^{\frac{n}{2}}G_b(u)
  G_b(q)G_{(bh^2)^{-1}}(x-v)dv\frac{dudq}{(2\pi
    h)^n}\biggr)^{\frac{2}{n}}
  \Biggr|\nonumber\\
  \leq Ch^{-2}(h^4a^2b+b^{-3/2})\leq C h^{-2+6/5}\label{eq:rhogamma2},
\end{eqnarray}
where $C$ may depend on $\tau$.

We now use that for all $x,y\in\R$ and all $n\geq3$ we have  
\begin{equation}
  \left||x_-|^{\frac{n}{2}}-|y_-|^{\frac{n}{2}} +
    \mfr{n}{2}|y_-|^{\frac{n}{2}-1}(x-y)\right|\leq 
  \left\{\begin{array}{cl}C|x-y|^{\frac{3}{2}},&n=3\\
      C(|x|^{\frac{n}{2}-2}+|y|^{\frac{n}{2}-2})|x-y|^2,&n\geq4
    \end{array}\right.\label{eq:3/2holder}
\end{equation}
where $C$ depends on $n$. This gives for $n=3$ (it is left to the
reader to write down the estimates for $n\geq 4$)
\begin{equation}\label{eq:3/2holder2}
\Biggl|\biggl|\biggl(V(v)+\eta_v(u,q)\biggr)_-\biggr|^{\frac{3}{2}}-
|V(v)_-|^{\frac{3}{2}}
+ \mfr{3}{2}|V(v)_-|^{\frac{1}{2}}\eta_v(u,q)\Biggr|
\leq C|\eta_v(u,q)|^{\frac{3}{2}}.
\end{equation}

It is now again crucial that 
$\int\eta_v(u,q)G_b(u)G_b(q)dudq=0$
and hence for $v\in\hbox{supp}(V)\subseteq B_{3/2}$
\begin{eqnarray}
  \biggl|\int\limits_{(1-h^2ab)u\in B_2-v}
  \eta_v(u,q)G_b(u)G_b(q)dudq\biggr|
  \leq Ce^{-b/5}\leq Ch^{6/5} . \label{eq:2ndorderexact}
\end{eqnarray}

Combining (\ref{eq:rhogamma2}), (\ref{eq:3/2holder2}),
(\ref{eq:2ndorderexact}), and
$|\eta_v(u,q)|\leq C(b^{-1}+|u|^2+|q|^2+h^2ab|u|)$ we obtain
\begin{eqnarray}
 \lefteqn{\biggl|\rho_\gamma(x)-(2\pi h)^{-3}\omega_3\int|V(v)_-|^{3/2}
 G_{(bh^2)^{-1}}(x-v)dv
  \biggr|}\nonumber\\
  &\leq& Ch^{-3}(e^{-b/5}+h^3a^{3/2}b^{3/4}+b^{-3/2}+h^{6/5})\leq
  Ch^{-3+6/5},\label{eq:rhogamma*}
\end{eqnarray}
where we have again removed the condition $(1-h^2ab)u\in B_2-v$ paying 
a price of $Ch^{-3}e^{-b/5}$. 

A simple Taylor expansion of $\phi^2$ gives 
$$
 \left|\phi(x)^2-\int\phi(v)^2G_{(bh^2)^{-1}}(x-v)dv\right|\leq C
 bh^2\leq Ch^{6/5},
$$
where we have again used that $\int vG_{(bh^2)^{-1}}(v)dv=0$.
This immediately gives (\ref{eq:rhogammaprop2}).

Finally, using again (\ref{eq:3/2holder}) we get
$$
  \left||V(x+v)_-|^{\frac{3}{2}}-|V(x)_-|^{\frac{3}{2}}
  + \mfr{3}{2}|V(x)_-|^{\frac{1}{2}}\nabla
  V(x)\cdot v\right|\leq C (|v|^{\frac{3}{2}}+|v|^2),
$$
and hence from (\ref{eq:rhogamma*})
\begin{eqnarray*}
 \biggl|\rho_\gamma(x)-(2\pi h)^{-3}\omega_n|V(x)_-|^{3/2}
  \biggr|\leq Ch^{-3}(h^{6/5}+(bh^2)^{3/4})
  \leq Ch^{-3+9/10} .
\end{eqnarray*}

We must now calculate
${\Tr}(\c\phi H\phi)={\Tr}(\c\phi (-h^2\Delta)\phi)+{\Tr}(\c\phi
V\phi)$ for $n\ge3$. {F}rom the argument leading to (\ref{eq:rhogammaprop2}) we
have 
\begin{equation}\label{eq:laplace0}
  (2\pi h)^n{\Tr}(\c\phi V\phi)\leq
  -\omega_n\int\phi(x)^2|V(x)_-|^{\frac{n}{2}+1}dx
  +Ch^{-n+6/5}.
\end{equation}
{F}rom Theorem \ref{trace formula} we have
\begin{eqnarray*}
  (2\pi h)^n{\Tr}(\c\phi(-h^2\Delta) \phi)
  &=&
  \int \,\chi({h}_{u,q}(v,p))\,G_b(u-v)G_b(q-p)
  \Big[E_2+\\&&(\phi(v+h^2ab(u-v))^2+E_1)
  (p+h^2ab(q-p))^2
    \Big]dudq dv dp,
\end{eqnarray*}
where $E_1, E_2$ are functions such that
$\|E_1\|_\infty,\|E_2\|_\infty\leq Ch^2b$. Since 
$$
 \int \,\chi({h}_{u,q}(v,p))\,G_b(u-v)G_b(q-p)
   (1+p^2)dudq dv dp\leq C,
$$
(note that it is important here that ${h}_{u,q}(v,p)=0$ unless 
$u\in B_2$) we get
\begin{eqnarray*}
  \lefteqn{(2\pi h)^n{\Tr}(\c\phi (-h^2\Delta)\phi)}&&\\
  &\le&
  \int \,\chi(h_{u,q}(v,p))\,G_b(u-v)G_b(q-p)\phi(v+h^2ab(u-v))^2\\&&\times\,
  (p+h^2ab(q-p))^2 dudq dv dp +Cbh^2.
\end{eqnarray*}
{F}rom (\ref{eq:happ}) we may now conclude that 
\begin{eqnarray}
  (2\pi h)^n{\Tr}(\c\phi ( -h^2\Delta)\phi)
  &\leq&\int \,\chi(\s(v,p)+\xi_v(u,q)-C|u|(b^{-1}+|u|^2))\,G_b(u)G_b(q)
  \nonumber\\ &&\times\, \phi(v+h^2abu)^2
  (p+h^2abq)^2dudq dv dp+Cbh^2\label{eq:laplace1}.
\end{eqnarray}
We now perform the $p$-integration in (\ref{eq:laplace1}) and arrive
at
\begin{eqnarray}
  (2\pi h)^n{\Tr}(\c\phi ( -h^2\Delta)\phi)
  &\leq&\frac{n}{n+2}\omega_n\int \, 
  \left|\left(V(v)+\xi_v(u,q)
    -C|u|(b^{-1}+|u|^2)\right)_-\right|^{\frac{n}{2}+1}
\nonumber\\&&\times\,G_b(u)G_b(q)
  \phi(v+h^2abu)^2
  dudq dv+Cbh^2,\label{eq:laplace2}
\end{eqnarray}
where we have used that the integral over the term containing 
$q\cdot p$ vanishes and the integral over the term containing
$(h^2abq)^2$ is bounded by $h^4a^2b\leq h^2b$.

We now expand the integrand in (\ref{eq:laplace2}) in the same way as
we did the integrand in (\ref{eq:lowerexpansion}). 
We finally obtain
$$
  (2\pi h)^n{\Tr}(\c\phi ( -h^2\Delta)\phi)
  \leq\frac{n}{n+2}\omega_n\int \, 
  \left|V(v)_-\right|^{\frac{n}{2}+1}
  \phi(v)^2dv+Ch^{6/5},
$$
which together with (\ref{eq:laplace0}) gives (\ref{eq:lemmaupper}).
\end{proof}

We shall need the generalization of Theorem~\ref{local semi-classics}
and Lemma~\ref{lm:upperbound} to a ball of arbitrary radius. 
We also require to know how the error term depends more explicitly on
the potential. 

\begin{cor}[Rescaled semi-classics] \label{corollary}
Let $n\geq 3$, $\phi\in C^{n+4}_0(\R^n)$ be supported in a ball
$B_\ell$ of radius 
$\ell>0$. Let $V\in C^3(\bar{B}_\ell)$ be a real potential. 
Let $H=-h^2\D +V$, $h>0$ and $\sigma(u,q) = q^2 + V(u)$. Then for all
$h>0$ and $f>0$ we have
\begin{equation} \label{eq:phiHphilf}
  \left|\Tr[\phi H\phi]_- - (2\pi h)^{-n}\int \phi(u)^2 \sigma(u,q)_-\, du dq \right|
  \le C h^{-n+6/5} f^{n+4/5}\ell^{n-6/5},
\end{equation}
where the constant $C$ depends only on 
\begin{equation}\label{eq:phivdependence}
  \sup_{|\a|\le n+4}\|\ell^{|\a|}\p^\a\phi\|_{\infty},\quad\hbox{ and }\quad
  \sup_{|\a|\le3}\|f^{-2}\ell^{|\a|}\p^\a V\|_{\infty}.
\end{equation}
Moreover, there exists a density matrix $\gamma$ such that 
\begin{equation}
\Tr[\phi H\phi\gamma]\leq (2\pi h)^{-n}\int \phi(u)^2 \sigma(u,q)_-\, du dq 
  +C h^{-n+6/5} f^{n+4/5}\ell^{n-6/5}\label{eq:gammaproplf}
\end{equation}
and such that its density  $\rho_\gamma(x)$ satisfies
\begin{equation}\label{eq:rhogammaproplf1}
\left|\rho_\gamma(x)-(2\pi h)^{-n}\omega_n
  \left|V(x)_-\right|^{n/2}\right|\leq Ch^{-n+9/10}f^{n-9/10}\ell^{-9/10},
\end{equation}
for (almost) all $x\in B_\ell$ and
\begin{equation}\label{eq:rhogammaproplf2}
\left|\int\phi(x)^2\rho_\gamma(x)dx-(2\pi h)^{-n}\omega_n\int\phi(x)^2
  \left|V(x)_-\right|^{n/2}dx\right|\leq Ch^{-n+6/5}f^{n-6/5}\ell^{n-6/5},
\end{equation}
where the constants $C>0$ in the above estimates again depend only on
the parameters in (\ref{eq:phivdependence}).
\end{cor}
\begin{proof} This is a simple rescaling argument. Introducing the
  unitary operator $(U\psi)(x)=\ell^{-n/2}\psi(\ell^{-1} x)$ we see
  that $\phi H\phi$ 
  is unitarily equivalent to the operator 
  $$
  U^*\phi H\phi U=f^2\phi_\ell(-h^2f^{-2}\ell^{-2}\Delta + V_{f,\ell})\phi_\ell,
  $$
  where $\phi_\ell(x)=\phi(\ell x)$, and $V_{f,\ell}(x)=f^{-2}V(\ell
  x)$.
  Thus 
  $$\Tr[\phi H\phi]_-=f^{2}
  \Tr[\phi_\ell(-h^2f^{-2}\ell^{-2}\Delta + V_{f,\ell})\phi_\ell]_-.
  $$
  Note that $\phi_\ell$ and $V_{f,\ell}$ are defined in a ball of
  radius $1$ and that for all $\a$
  $$
  \|\p^\a\phi_\ell\|_\infty 
  =\|\ell^{|\a|}\p^\a\phi\|_\infty,\quad
  \hbox{ and }\quad
  \|\p^\a V_{f,\ell}\|_{\infty}=
  \|f^{-2}\ell^{|\a|}\p^\a V\|_{\infty}.
  $$
  It follows from Theorem~\ref{local semi-classics} that
  \begin{equation} 
    \left|\Tr[\phi H\phi]_- 
      - (2\pi hf^{-1}\ell^{-1})^{-n}\int \phi_\ell(u)^2
      f^2\sigma_{f,\ell}(u,q)_-\, 
      du dq \right|
    \le C f^2 (hf^{-1}\ell^{-1})^{-n+6/5},
  \end{equation}
  where $\sigma_{f,\ell}(u,q)= q^2-V_{f,\ell}(u)$ and where the
  constant $C$ only depends on the parameters in
  (\ref{eq:phivdependence}). 
  A simple change of variables gives 
  $$
  (2\pi hf^{-1}\ell^{-1})^{-n}\int \phi_\ell(u)^2
  f^2\sigma_{f,\ell}(u,q)_-\, 
  du dq= (2\pi h)^{-n}\int \phi(u)^2
  \sigma(u,q)_-\, 
  du dq.
  $$
  Thus (\ref{eq:phiHphilf}) follows.

  To find the appropriate density matrix $\gamma$. We begin with the
  corresponding density matrix $\gamma_{f,\ell}$ for
  $\phi_\ell(-h^2f^{-2}\ell^{-2}\Delta + V_{f,\ell})\phi_\ell$, i.e.  the
  density matrix, which according to Lemma~\ref{lm:upperbound}
  satisfies the three estimates
  \beax
  \Tr\left[\phi_\ell(-h^2f^{-2}\ell^{-2}\Delta + V_{f,\ell})
    \phi_\ell\gamma_{f,\ell}\right]&\leq& 
    (2\pi hf^{-1}\ell^{-1})^{-n} \int\phi^2_\ell(u) \sigma_{f,\ell}(u,q)_- \,
    du dq 
    \\&&+C(hf^{-1}\ell^{-1})^{-n+6/5},
    \eeax
  $$
  \left|\rho_{\gamma_{f,\ell}}(x)-(2\pi hf^{-1}\ell^{-1})^{-n}
    \omega_n|V_{f,\ell}(x)_-|^{n/2}\right|
  \leq C(hf^{-1}\ell^{-1})^{-n+9/10} ,
  $$$$
  \left|\int\phi_\ell^2\rho_{\gamma_{f,\ell}}
    -(2\pi hf^{-1}\ell^{-1})^{-n}\omega_n\int\phi_\ell(x)^2
    |V_{f,\ell}(x)_-|^{n/2}dx\right|\leq C(hf^{-1}\ell^{-1})^{-n+6/5}.
  $$
  The density matrix $\gamma=U\gamma_{f,\ell}U^*$ whose density is 
  $\rho_\gamma(x)=\ell^{-n}\rho_{\gamma_{f,\ell}}(x/\ell)$ then satisfies 
  the properties (\ref{eq:gammaproplf}--\ref{eq:rhogammaproplf2}).
\end{proof}

\section{Semiclassics for the Thomas-Fermi potential}\label{sec:sctf}
We shall consider the semiclassical approximation for a Schr\"odinger
operator with the Thomas-Fermi potential $\V({\mathbf z},{\mathbf
  r},x)$, i.e., $-h^2\Delta-\V$. We shall throughout this section
simply write $\V(x)$ instead of $\V({\mathbf z},{\mathbf r},x)$.
Recall that $\V(x)>0$.

The main result we shall prove here is the Scott correction to the
semiclassical expansion for this potential.
\begin{thm}[Scott corrected semiclassics]\label{TF} 
  For all $h>0$ and all $r_1,\ldots,r_M\in\R^3$ with
  $\min_{k\ne m}|r_m-r_k|>r_0>0$ we have 
  \begin{equation}\label{eq:main1}
    \left|\Tr[-h^2\D - V^{\rm TF}]_- - (2\pi
      h)^{-3} \int (p^2 - V^{\rm TF}(u))_- \,du dp-
      \frac{1}{8h^2} \sum_{k=1}^M z_k^2\right| 
    \le C h^{-2+\frac{1}{10}},
  \end{equation}
  where $C>0$ depends only on $z_1,\ldots,z_M$, $M$, and $r_0$.
  Moreover, we can find a density matrix $\gamma$ such that 
  \begin{equation}\label{eq:maingamma1}
    \Tr \left[(-h^2\Delta - V^{\rm TF})\gamma\right]
    \leq \Tr \left[-h^2\Delta - V^{\rm TF}\right]_-+C h^{-2+1/10},
  \end{equation}
  and such that 
  \begin{equation}\label{eq:maingamma2}
    D\left(\rho_\gamma-\frac{1}{6\pi^2h^3}(V^{\rm
        TF})^{3/2}\right)\leq
    Ch^{-5+4/5}
  \end{equation}
  and
  \begin{equation}\label{eq:maingamma3}
    \int \rho_\gamma\leq \frac{1}{6\pi^2h^3}\int V^{\rm
      TF}(x)^{3/2}dx+C h^{-2+1/5},
  \end{equation}
  with $C$ depending on the same parameters as before.
\end{thm}
Note that if we choose $h=2^{-1/2}$ we have from
(\ref{eq:tfeqgeneral}) that
$(6\pi^2h^3)^{-1}(\V)^{3/2}=\rho^{\rm TF}/2$. The factor $1/2$ on the
right is due to the fact that we have not included  spin degeneracy in
Theorem~\ref{TF}.

In order to prove this theorem we shall compare with semiclassics for 
hydrogen like atoms. 
\begin{lemma}[Hydrogen comparison]\label{lm:hydrogencom}
For all $h>0$ and all $r_1,\ldots,r_M\in\R^3$ with
$\min_{k\ne m}|r_m-r_k|>r_0>0$ we have
\begin{eqnarray}
  \lefteqn{\Biggl|\Tr \left[-h^2\Delta-V^{\rm TF}(\x)\right]_-
    -(2\pi h)^{-3}\int \left(p^2-V^{\rm TF}(u)\right)_-dudp}&&\nonumber\\
  &&-\sum_{k=1}^M\left(\Tr
    \Bigl[-h^2\Delta-\frac{z_k}{|\x-r_k|}+1\Bigr]_-
    -(2\pi h)^{-3}\int
    \Bigl(p^2-\frac{z_k}{|u-r_k|}+1\Bigr)_-dudp\right)
  \Biggr| \nonumber\\
  &\leq& C h^{-2+1/10},\label{eq:hydcom}
\end{eqnarray}
where $C>0$ depends only on $z_1,\ldots,z_M$, $M$ and $r_0$.
\end{lemma}
The first estimate in Theorem~\ref{TF} follows from
Lemma~\ref{lm:hydrogencom} combined with the 
exact calculations for hydrogen
$$
\Tr\Bigl[-h^2\Delta-\frac{z_k}{|\x-r_k|}+1\Bigr]_-
=\sum_{1\leq n\leq z_k/(2h)}(-\frac{z_k^2}{4h^2}+n^2)
=-\frac{z_k^3}{12 h^3}+\frac{z_k^2}{8h^2}+{\cal O}(h^{-1})
$$
and
$$
 (2\pi h)^{-3}\int \Bigl(p^2-\frac{z_k}{|u-r_k|}+1\Bigr)_-dudp
 =-\frac{32\pi^2 z_k^3}{15(2\pi h)^{3}}\frac{\Gamma(7/2)\Gamma(1/2)}{\Gamma(4)} 
 =-\frac{z_k^3}{12 h^3}.
$$

Before giving the proof of Lemma~\ref{lm:hydrogencom} we introduce the 
function
\begin{equation}\label{eq:ldefinition}
  \ell(x)=\mfr{1}{2}\Bigl(1+\sum_{k=1}^M(|x-r_k|^2+\ell_0^2)^{-1/2}\Bigr)^{-1}
\end{equation}
where $0<\ell_0<1$ is a parameter that we shall choose explicitly in 
(\ref{eq:ell0choice}) below. Note
that $\ell$ is a smooth function with 
$$
0<\ell(x)<1,\quad\hbox{and}\quad
\|\nabla\ell(x)\|_\infty<1.
$$
Note also that in terms of the function $d(x)$ from (\ref{ddefinition})
we have  
\begin{equation}\label{eq:elldcom}
 \mfr{1}{2}(1+M)^{-1}\ell_0\leq
 \mfr{1}{2}(1+M(d(x)^2+\ell_0^2)^{-1/2})^{-1}
 \leq\ell(x)\leq\mfr{1}{2}(d(x)^2+\ell_0^2)^{1/2}.
\end{equation}
Note in particular that we have 
\begin{equation}\label{eq:elldcom2}
\ell(x)\geq\mfr{1}{2}(1+M)^{-1}\min\{d(x),1\}.
\end{equation}
We fix a localization function $\phi\in C_0^\infty(\R^3)$ with support
in $\{|x|<1\}$ and such that $\int\phi(x)^2dx=1$.
According to Theorem~\ref{partition} we can find a corresponding
family of functions $\phi_u\in C_0^\infty(\R^3)$, $u\in\R^3$, 
where $\phi_u$ is supported in the ball $\{|x-u|<\ell(u)\}$ with the
properties that 
\begin{equation}\label{eq:phiuprop}
  \int\phi_u(x)^2\ell(u)^{-3}du=1\quad\hbox{and}\quad
  \|\partial^\alpha\phi_u\|_\infty\leq C\ell(u)^{-|\alpha|},
\end{equation} 
for all multi-indices $\alpha$, where $C>0$
depends only on $\alpha$ and $\phi$.

Moreover, from (\ref{eq:tfdf}) in Theorem~\ref{thm:tfestimate} we know that for all $u\in
\R^n$ with $d(u)>2\ell_0$ the  TF-potential $V^{\rm TF}$ satisfies 
\begin{equation}\label{eq:tflf}
  \sup_{|x-u|<\ell(u)}|\p^\alpha V^{\rm TF}(x)|\leq Cf(u)^2\ell(u)^{-|\alpha|},
\end{equation}
where $C>0$ depends only on $\alpha$, $z_1,\ldots,z_M$, and $M$.
We have here used the fact that if $d(u)>2\ell_0$ then
$\ell(u)\leq \sqrt{5}d(u)/4$ and hence for all $x$ with $|x-u|<\ell(u)$ we have 
(note that $d(u)\le d(x)+|x-u|$ and $\sqrt{5}/4<1$)
$$
 \ell(u)<C d(x)\quad\hbox{and}\quad f(x)\leq C f(u).
$$

\begin{proof}[Proof of Lemma~\ref{lm:hydrogencom}]
We note first that we may if necessary assume that $h$ is smaller than
some constant depending only on the parameters $z_1,\ldots,z_M$, $M$,
$r_0$. This follows from the Lieb-Thirring inequality (\ref{LT}) and
the estimate on $V^{\rm TF}$ given in (\ref{eq:tfdf}) for $\alpha=0$.

In order to control the region far away from all the nuclei we
introduce localization functions $\theta_-,\theta_+\in C^\infty(\R)$
such that
\begin{enumerate}
\item $\theta_-^2+\theta_+^2=1$, 
\item $\theta_-(t)=1$ if $t<1$ and $\theta_-(t)=0$ if $t>2$. 
\end{enumerate}
Let  
\begin{equation}\label{eq:Rchoice}
  R=h^{-1/2}
\end{equation}
and define 
$\Phi_\pm(x)=\theta_\pm(d(x)/R)$. Then 
$\Phi_-^2+\Phi_+^2=1$. Denote ${\cal
  I}=(\nabla\Phi_-)^2+(\nabla\Phi_+)^2$. 
Then ${\cal I}$ is supported on a set whose volume is bounded by $CR^3$
(where as before $C$ depends on $M$) and 
$$
\|{\cal I}\|_\infty\leq CR^{-2}.
$$

Using the  IMS-formula (\ref{eq:IMS}) we find that 
$$
  -h^2\Delta-V^{\rm TF}=\Phi_-(-h^2\Delta-V^{\rm TF}-h^2{\cal I})\Phi_-
  +\Phi_+(-h^2\Delta-V^{\rm TF}-h^2{\cal I})\Phi_+
$$
{F}rom the Lieb-Thirring inequality the estimates on ${\cal I}$ and the bound 
$V^{\rm TF}(x)\leq Cd(x)^{-4}$ (see (\ref{eq:tfdf}) with $\alpha=0$) we find 
$$
\Tr[-h^2\Delta-V^{\rm TF}]_-\geq \Tr[\Phi_-(-h^2\Delta-V^{\rm TF}-h^2{\cal
  I})\Phi_-]_- -C(h^{-3}R^{-7} + h^{2} R^{-2}).
$$ 
On the support of $\Phi_-$ we now use the localization functions
$\phi_u$. Again using the IMS formula (\ref{eq:IMS}) we obtain from
(\ref{eq:phiuprop}) that 
\begin{eqnarray*}
  \lefteqn{\Phi_-\left(-h^2\Delta-V^{\rm TF}-h^2{\cal
        I}\right)\Phi_-}&&\\
  &\geq& 
  \int \Phi_-\phi_u\left(-h^2\Delta-V^{\rm TF}-Ch^2(\ell(u)^{-2}
    +R^{-2})\right)\phi_u\Phi_-\ell(u)^{-3}du.
\end{eqnarray*}
We have here used that if the supports of $\phi_u$ and $\phi_{u'}$
overlap then $|u-u'|\leq\ell(u)+\ell(u')$ and thus
$$
 \ell(u)\leq\ell(u')+\|\nabla\ell\|_\infty(\ell(u)+\ell(u')).
$$
Therefore, since $\|\nabla\ell\|_\infty<1$, we have that  $\ell(u)\leq
C\ell(u')$ and thus $\ell(u')^{-2}\leq C\ell(u)^{-2}$.

{F}rom the variational principle we now get
\begin{eqnarray}
  \lefteqn{\Tr[-h^2\Delta-V^{\rm TF}]_-}&&\label{eq:tfphilow}\\\nonumber
  &\geq&\int_{d(u)<2R+1}\Tr[\phi_u\left(-h^2\Delta-V^{\rm
      TF}-Ch^2\ell(u)^{-2}\right)\phi_u]_-  
  \ell(u)^{-3}du\\&&-C(h^{-3}R^{-7} + h^2R^{-2}),\nonumber
\end{eqnarray}
where we have restricted the integral
according to the support of $\Phi_-$ and $\phi_u $ and used that, since
we may assume that $h$ is so small that $R>C$, 
then $\ell(u)^{-2}\geq CR^{-2}$.  Note that there is no need to
write $\Phi_-$ on the right, since in general
$\Tr(\Phi_-A\Phi_-)_-\geq \Tr A_-$ for any selfadjoint operator $A$.

In a very similar manner we get corresponding estimates for the
hydrogenic operators. In particular, if we choose $h$ so small that
$R>\max_k\{z_k\}$ then on the support of $\Phi_+$ we have
$-z_k|x-r_k|^{-1}+1\geq0$. Thus we have
\begin{eqnarray}
  \lefteqn{\Tr\Bigl[-h^2\Delta-\frac{z_k}{|\hat{x}-r_k|}+1\Bigr]_-}&&
  \label{eq:hydphilow}\\
  &\geq&\int_{d(u)<2R+1}\Tr\Bigl[\phi_u\Bigl(-h^2\Delta-\frac{z_k}{|\hat{x}-r_k|}+1
  -Ch^2\ell(u)^{-2}\Bigr)\phi_u\Bigr]_- \ell(u)^{-3}du\nonumber\\&&-Ch^{2} R^{-2}.\nonumber
\end{eqnarray}

We shall now get upper bounds similar to (\ref{eq:tfphilow}) and
(\ref{eq:hydphilow}).  If we again denote by $\chi$ the characteristic
function of the interval $(-\infty,0]$ we see from (\ref{eq:phiuprop})
that
$$
\gamma=\int_{d(u)<2R+1}\phi_u\chi\left(\phi_u(-h^2\Delta-V^{\rm TF})\phi_u\right)\phi_u\ell(u)^{-3}du
$$ 
defines a density matrix. If we use it as a trial density matrix to get 
an upper bound we obtain
\begin{eqnarray}
  \Tr[-h^2\Delta-V^{\rm TF}]_-&\leq& \Tr[(-h^2\Delta-V^{\rm TF})\gamma]\nonumber\\
  &=&\int\limits_{d(u)<2R+1}\Tr[\phi_u\left(-h^2\Delta-V^{\rm
      TF}\right)\phi_u]_-  \ell(u)^{-3}du.\label{eq:tfphiup}
\end{eqnarray}
Similarly, 
\begin{eqnarray}
  \Tr\Bigl[-h^2\Delta-\frac{z_k}{|\hat{x}-r_k|}+1\Bigr]_-\leq
  \int\limits_{d(u)<2R+1}\!\!\!\!\Tr\Bigl[\phi_u\Bigl(-h^2\Delta-\frac{z_k}{|\hat{x}-r_k|}+1\Bigr)\phi_u\Bigr]_-  
  \ell(u)^{-3}du.\label{eq:hydphiup}
\end{eqnarray}

We now introduce the quantities
\begin{eqnarray*}
  D_+(u)&:=&\Tr[\phi_u\left(-h^2\Delta-V^{\rm
      TF}-Ch^2\ell(u)^{-2}\right)\phi_u]_-\nonumber\\&&-
  \sum_{k=1}^M\Tr\Bigl[\phi_u\Bigl(-h^2\Delta-\frac{z_k}{|\hat{x}-r_k|}+1\Bigr)\phi_u\Bigr]_- ,
   \\
D_-(u)&:=&\sum_{k=1}^M\Tr\Bigl[\phi_u\Bigl(-h^2\Delta
  -\frac{z_k}{|\hat{x}-r_k|}+1-Ch^2\ell(u)^{-2}\Bigr)\phi_u\Bigr]_-
  \nonumber\\&&-
  \Tr[\phi_u(-h^2\Delta-V^{\rm
    TF})\phi_u]_- ,\\
\noalign{and}
D_{\rm SC}(u)&:=&(2\pi h)^{-3}\int\phi_u(x)^2(p^2-\V(x))_-dpdx\nonumber\\&&-
(2\pi h)^{-3}\sum_{k=1}^M\int\phi_u(x)^2\Bigl(p^2-\frac{z_k}{|x-r_k|}+1\Bigr)_-dpdx .
\end{eqnarray*}
Then from (\ref{eq:tfphilow}), (\ref{eq:hydphilow}),(\ref{eq:tfphiup}),
 and (\ref{eq:hydphiup}) we have 
\begin{eqnarray}
  \Tr[-h^2\Delta-\V]_- 
    - \sum_{k=1}^M\Tr\Bigl[-h^2\Delta-\frac{z_k}{|\hat{x}-r_k|}+1\Bigr]_-
  &\geq&\int_{d(u)<2R+1} D_+(u)\ell(u)^{-3}du\nonumber\\&&
  -C(h^2R^{-2}+h^{-3}R^{-7})\label{eq:Dintegral}
\end{eqnarray}
and
\begin{eqnarray}
  \sum_{k=1}^M\Tr\Bigl[-h^2\Delta-\frac{z_k}{|\hat{x}-r_k|}+1\Bigr]_--\Tr[-h^2\Delta-\V]_-
  &\geq&\int_{d(u)<2R+1}  D_-(u)\ell(u)^{-3}du\nonumber\\
  &&-Ch^2R^{-2},\label{eq:D-integral}
\end{eqnarray}
and from (\ref{eq:phiuprop}) 
\begin{eqnarray}
  (2\pi h)^{-3}\int(p^2-\V(x))_-dpdx-
    (2\pi h)^{-3}\sum_{k=1}^M\int\Bigl(p^2-\frac{z_k}{|x-r_k|}+1\Bigr)_-dpdx
  \nonumber\\=
  \int D_{\rm SC}(u)\ell(u)^{-3}du.\label{eq:Dscintegral}
\end{eqnarray}
The same estimates  which led to (\ref{eq:tfphilow}) and
(\ref{eq:hydphilow}) give
\begin{eqnarray}\label{eq:DscR}
  \left|\int D_{\rm SC}(u)\ell(u)^{-3}du-\int_{d(u)<2R+1} D_{\rm
      SC}(u)\ell(u)^{-3}du\right|
  \leq Ch^{-3}R^{-7}.
\end{eqnarray}

We shall prove the lemma by establishing lower bounds on 
$D_+(u)-D_{\rm SC}(u)$ and $D_-(u)+D_{\rm SC}(u)$.

We consider first $u$ with $d(u)\leq 2\ell_0$, where
$\ell_0$ is the parameter that occurs in the definition (\ref{eq:ldefinition})
of $\ell$. We choose 
\begin{equation}\label{eq:ell0choice}
  \ell_0=h,
\end{equation}
where we assume that $h$ is small enough to ensure that $\ell_0<1$.
In fact, we may also assume that $\ell_0<r_0/8$. If $k$ is such that
$d(u)=|u-r_k|$ we get from (\ref{eq:elldcom}) that 
$$
|u-r_k|+\ell(u)=d(u)+\ell(u)\leq
d(u)+\mfr{1}{2}(d(u)^2+\ell_0^2)^{1/2} < 4\ell_0<r_0/2.
$$
Thus for all $x$ with $|x-u|<\ell(u)$, i.e., for all $x$ in the
support of $\phi_u$ we must have that $|x-r_k|<r_0/2$
and hence $|x-r_j|\geq r_0/2$. Thus
$d(x)=|x-r_k|$ and  of the nuclear positions $r_1,\ldots,r_M$ only
$r_k$ may be contained in the support of $\phi_u$.
Since the function $W_k(x)=V^{{\rm TF}}(x)-z_k|x-r_k|^{-1}$ satisfies
the estimate (\ref{eq:westimate}) on the support of $\phi_u$
we have for $0<\e<1/2$ that
\begin{eqnarray*}
  \lefteqn{\Tr\Big[\phi_u\big(-h^2\D
    -\V-Ch^2\ell(u)^{-2}\big)\phi_u\Big]_-}&&\\
  &\geq&
  \Tr\Big[\phi_u\Big(-(1-\e)h^2\D -(1-\e)\frac{z_k}{|\hat{x}-r_k|} +
  (1-\e)\Big)\phi_u\Big]_-
  \\
  &&+ \,\Tr\Big[\phi_u\Big(-\e h^2\D - \e \frac{z_k}{|\hat{x}-r_k|} -(1-\e) -W_k(x) -
  Ch^2\ell(u)^{-2}\Big)\phi_u\Big]_-\\
  &\geq&
  \Tr\Big[\phi_u\Big(-h^2\D -\frac{z_k}{|\hat{x}-r_k|} + 1\Big)\phi_u\Big]_-
  \\
  &&-Ch^{-3}[\e \ell(u)^{1/2}+\e^{-3/2}\ell(u)^3(1+r_0^{-5/2})
  +h^5\e^{-3/2}\ell(u)^{-2}],
\end{eqnarray*}
where in the last line we have used the Lieb-Thirring inequality
(\ref{LT}) and the fact that $\Tr[\ldots]_-\leq0$. 
We therefore have that 
\begin{equation}\label{eq:Dusmall}
  D_+(u)\geq -Ch^{-3}[\e \ell(u)^{1/2}+\e^{-3/2}\ell(u)^3
  +h^5\e^{-3/2}\ell(u)^{-2}].
\end{equation}

The quantity $D_-(u)$ is estimated in essentially the same way. We 
get 
\begin{eqnarray*}
  \lefteqn{\Tr\Bigl[\phi_u\Bigl(-h^2\Delta
    -\frac{z_k}{|\hat{x}-r_k|}+1-Ch^2\ell(u)^{-2}\Bigr)\phi_u\Bigr]_-}&&\\
  &\geq& \Tr\Bigl[\phi_u\Bigl(-(1-\e)h^2\Delta
  -(1-\e)\V\Big)\phi_u\Big]_-\\&&+\Tr\Bigl[\phi_u\Bigl(-\e h^2\Delta-\e\frac{z_k}{|\hat{x}-r_k|}
  +(1-\e)W_k-Ch^2\ell(u)^{-2}\Bigr)\phi_u\Bigr]_-
  \\&\geq& \Tr\Bigl[\phi_u\Bigl(-h^2\Delta
  -\V\Big)\phi_u\Bigr]_-\\&&
  -Ch^{-3}[\e \ell(u)^{1/2}+\e^{-3/2}\ell(u)^3
  +h^5\e^{-3/2}\ell(u)^{-2}],
\end{eqnarray*}
and again by the Lieb-Thirring inequality
\begin{eqnarray*}
\sum_{j, j\ne k}\Tr\Bigl[\phi_u\Bigl(-h^2\Delta
-\frac{z_k}{|\hat{x}-r_j|}+1-Ch^2\ell(u)^{-2}\Bigr)\phi_u\Bigr]_-
\geq -Ch^{-3}[\ell(u)^3 +h^5\ell(u)^{-2}].
\end{eqnarray*}
Therefore $D_-(u)$ satisfies an estimate similar to (\ref{eq:Dusmall}).
Since $d(u)\leq 2\ell_0$ and $\ell_0<1$ 
we have that  $C^{-1}\ell_0\leq \ell(u)\leq C\ell_0$.
Hence we can replace $\ell(u)$ by $\ell_0$ in the above estimates if we 
change the constant $C$. If we now choose $\e=\ell_0^{-1} h^2$ 
(with the choice (\ref{eq:ell0choice}) we may assume that
$h$ is so small that, indeed, $\e<1/2$) we get
\begin{equation}\label{eq:D-Dusmall}
  D_+(u),D_-(u)\geq -Ch^{-3}[h^2\ell_0^{-1/2}+
 h^{-3}\ell_0^{9/2}].
\end{equation}
By an identical argument using $(x+y)_-\geq x_-+y_-$ we get that
for $u$ with $d(u)\leq 2\ell_0$ we have  
\begin{equation}
  |D_{\rm SC}(u)|\leq Ch^{-3}[h^2\ell_0^{-1/2}+
  h^{-3}\ell_0^{9/2}].
  \label{eq:usmallsc}
\end{equation}

We next consider  $u$ such that $2\ell_0<d(u)\leq 2R+1$. 
We choose again $0<\e<1/2$ (not necessarily the same as before) and
use the Lieb-Thirring inequality to arrive at 
\begin{eqnarray}
  D_+(u)&\geq&\Tr[\phi_u\left(-(1-\e)h^2\Delta-\V\right)\phi_u]_-\nonumber\\
  &&-\sum_{k=1}^M\Tr\Bigl[\phi_u\Bigl(-h^2\Delta
  -\frac{z_k}{|\hat{x}-r_k|}+1\Bigr)\phi_u\Bigr]_-
  -C\e^{-3/2}h^2\ell(u)^{-2}.
  \label{eq:dulargelower}
\end{eqnarray}
Since
$\V$ satisfies the estimate (\ref{eq:tflf}) on the 
ball $\{x\ |\ |x-u|<\ell(u)\}$ and $\phi_u$ satisfies
(\ref{eq:phiuprop}) we may use Corollary~\ref{corollary}
to conclude that 
\begin{eqnarray*}
  \left|\Tr[\phi_u\left(-(1-\e)h^2\Delta-\V\right)\phi_u]_-
      -(2\pi (1-\e)^{1/2}h)^{-3}\int\phi_u^2(x)
      (p^2-\V(x))_-dxdp\right|
    \\
  \leq C h^{-3}h^{6/5}f(u)^{19/5}\ell(u)^{9/5},
\end{eqnarray*}
where $C$ depends only on $z_1,\ldots,z_M$ and $M$. Hence again using 
(\ref{eq:tflf}) we obtain 
\begin{eqnarray}
  \Tr[\phi_u\left(-(1-\e)h^2\Delta-\V\right)\phi_u]_-&\geq&
      (2\pi h)^{-3}\int\phi_u^2(x) (p^2-\V(x))_-dxdp
  \label{eq:dulargesc1}\\
  &&-Ch^{-3}(\e f(u)^5\ell(u)^3  +h^{6/5}f(u)^{19/5}\ell(u)^{9/5}).\nonumber
\end{eqnarray}
If $d(u)>\max_k\{z_k\}+1$ then we have that for all $k=1,\ldots,M$ that 
$$
\Tr\Bigl[\phi_u\Bigl(-h^2\Delta-\frac{z_k}{|\hat{x}-r_k|}+1\Bigr)\phi_u\Bigr]_-=0
\quad\hbox{and}\quad
\int\phi_u(x)^2\Bigl(p^2-\frac{z_k}{|x-r_k|}+1\Bigr)_-dpdx=0.
$$
On the other hand, if $2\ell_0<d(u)<\max_k\{z_k\}+1$ then for all $k=1,\ldots,M$ 
the potential $z_k|x-r_k|^{-1}-1$ satisfies a bound similar to 
(\ref{eq:tflf}) and we may again use Corollary~\ref{corollary} to
conclude that for all $u$ with $d(u)>2\ell_0$ we have 
\begin{eqnarray}
  \Bigl|\Tr\Bigl[\phi_u\Bigl(-h^2\Delta
  -\frac{z_k}{|\hat{x}-r_k|}+1\Bigr)\phi_u\Bigr]_--
  (2\pi h)^{-3}\int\phi_u^2(x) 
  \Bigl(p^2-\frac{z_k}{|x-r_k|}+1\Bigr)_-dxdp\Bigr|\nonumber\\
  \leq C h^{-3}h^{6/5}f(u)^{19/5}\ell(u)^{9/5}.\label{eq:dulargesc2}
\end{eqnarray}
Hence from (\ref{eq:dulargelower}), (\ref{eq:dulargesc1}), 
and (\ref{eq:dulargesc2})
we have for all $u$ with $2\ell_0<d(u)\leq 2R+1$ that 
\begin{eqnarray}
  D_+(u)-D_{\rm SC}(u)\nonumber
  &\geq&-Ch^{-3}(\e^{-3/2}h^5\ell(u)^{-2}+\e f(u)^5\ell(u)^3
  +h^{6/5}f(u)^{19/5}\ell(u)^{9/5})\nonumber
  \\&=&-Ch^{-3}(h^2f(u)^3\ell(u)
  +h^{6/5}f(u)^{19/5}\ell(u)^{9/5}),\label{eq:D+lowerfinal}
\end{eqnarray}
where we have chosen $\e=ch^2\ell(u)^{-2}f(u)^{-2}$. Note that from  
the property (\ref{eq:elldcom2}) of $\ell$  and the definition
(\ref{fdefinition}) of $f$ we have
$$\e\leq cCh^2\max\{d(u)^{-1},d(u)^4\}\leq cCh^2 \max\{\ell_0^{-1},(2R+1)^4\}.$$ 
We see that with the choice of $R$
in (\ref{eq:Rchoice}) and of $\ell_0$ in (\ref{eq:ell0choice}) 
we may assume that $h$ and $c$ are chosen small enough so  
that $\e<1/2$.

In a completely similar way we get
for all $u$ with $2\ell_0<d(u)\leq 2R+1$ that 
\begin{eqnarray}\label{eq:D-lowerfinal}
D_-(u)+D_{\rm SC}(u)
  &\geq&-Ch^{-3}(h^2f(u)^3\ell(u)
  +h^{6/5}f(u)^{19/5}\ell(u)^{9/5}).
\end{eqnarray}

If we now combine
(\ref{eq:D-Dusmall}),(\ref{eq:usmallsc}),
(\ref{eq:D+lowerfinal}), and (\ref{eq:D-lowerfinal}) we obtain
\begin{eqnarray}
  \lefteqn{\int\limits_{d(u)\leq 2R+1} [D_\pm(u)\mp D_{\rm SC}(u)
    ]\ell(u)^{-3} du}&&\nonumber
  \\&\geq&
  -Ch^{-3}[h^2\ell_0^{-1/2}+
 h^{-3}\ell_0^{9/2}]\nonumber\\
 &&-Ch^{-3}\!\!\!\!\!\!\!\!\!\!
 \int\limits_{2\ell_0\leq d(u)\leq 2R+1}\!\!\!\!\! (h^2f(u)^3\ell(u)^{-2}
  +h^{6/5}f(u)^{19/5}\ell(u)^{-6/5})du, \label{eq:DDsccom}
\end{eqnarray}
where we have used that the volume of the set of $u$ for which
$d(u)\leq 2\ell_0$ is bounded by $C\ell_0^3$ and that from (\ref{eq:elldcom}),
$\ell(u)\geq C^{-1}\ell_0$. Using again
the property (\ref{eq:elldcom2}) of $\ell$  and the definition
(\ref{fdefinition}) of $f$ we see that the last integral
in (\ref{eq:DDsccom}) is bounded by 
\begin{eqnarray*}
  &&C h^{-3}\!\!\!\!\!\!\!\!\!\!
  \int\limits_{2\ell_0\leq d(u)\leq
    2R+1}\!\!\!\!\!h^2\min\{d(u)^{-7/2},d(u)^{-6}\}
  +h^{6/5}\min\{d(u)^{-31/10},d(u)^{-38/5}\}du
  \\
  &\leq&Ch^{-3}
  \int\limits_{2\ell_0\leq |u|}h^2\min\{|u|^{-7/2},|u|^{-6}\}
  +h^{6/5}\min\{|u|^{-31/10},|u|^{-38/5}\}du\\
  &\leq&C h^{-1}\ell_0^{-1/2}+Ch^{-9/5}\ell_0^{-1/10}.
\end{eqnarray*}
Thus if we combine (\ref{eq:Dintegral}), (\ref{eq:D-integral}),
(\ref{eq:Dscintegral}) ,(\ref{eq:DscR}), and (\ref{eq:DDsccom})
we see that the left side of the main inequality (\ref{eq:hydcom}) is
bounded by
\begin{eqnarray*}
  C (h^{-1}\ell_0^{-1/2}+h^{-9/5}\ell_0^{-1/10}+
  h^{-6}\ell_0^{9/2}
  +h^2R^{-2}+Ch^{-3}R^{-7})\\
  \leq C(h^{-3/2}+h^{-19/10}+h^{-3/2}+h^3+h^{1/2})\leq Ch^{-2+1/10}.
\end{eqnarray*}
Note that the choice of $\ell_0$ has not been optimized. 
\end{proof}
\begin{proof}[Proof of Theorem~\ref{TF}]
As mentioned just after the statement of Lemma~\ref{lm:hydrogencom}
the first estimate in the theorem is a consequence of the lemma.  
It remains to prove the existence of a density matrix $\gamma$ with the stated 
properties. 

We note first that we may as before, if necessary, assume
that $h$ is smaller than
some constant depending only on the parameters $z_1,\ldots,z_M$, $M$, and
$r_0$. Otherwise we simply choose $\gamma=0$. That this is an
acceptable choice follows from the Lieb-Thirring inequality (\ref{LT}) 
and the fact that the estimate (\ref{eq:tfdf}) for $\alpha=0$ implies 
that $D\left(\left(V^{\rm TF}\right)^{3/2}\right)\leq C$. 

To construct $\gamma$ we shall again use the localization family $\phi_u$ with
the properties given in (\ref{eq:phiuprop}). As in the previous lemma 
we shall choose $R=h^{-1/2}$ and $\ell_0=h$ (although it is not a requirement that
they should be as before). We shall choose $\gamma$ of the form
\begin{equation}\label{eq:gammachoice}
  \gamma=\int_{d(u)<R} \phi_u\gamma_u\phi_u\ell(u)^{-3}du,
\end{equation}
where $\gamma_u$ is a family of density matrices which we shall now
choose. 
Note that the first condition in (\ref{eq:phiuprop}) implies
that $\gamma$ is then a density matrix.  

If $2\ell_0<d(u)$ it follows from (\ref{eq:phiuprop}),
(\ref{eq:tflf}), and Corollary~\ref{corollary}  
that we may choose  $\gamma_u$ such that (\ref{eq:gammaproplf}),
(\ref{eq:rhogammaproplf1}), and  (\ref{eq:rhogammaproplf2}) hold 
when $V=V^{\rm TF}$, $\phi=\phi_u$, $\ell=\ell(u)$ and $f=f(u)$. 

If $d(u)\leq 2\ell_0$ we simply choose  
$$
  \gamma_u=
  \chi\left(\phi_u\left(-h^2\Delta-V^{\rm TF}\right)\phi_u\right),
  $$
  where $\chi$ is again the characteristic function of the interval
  $(-\infty,0]$. I.e., $\gamma_u$ is the projection onto the
  non-positive spectrum of $\phi_u\left(-h^2\Delta-V^{\rm
      TF}\right)\phi_u$. Here we are considering $\phi_u$ as a
  multiplication operator.

We shall first prove that for  $d(u)\leq 2\ell_0$ we have 
\begin{equation}\label{eq:rhogammu5/3}
  \int \left(\phi_u^2\rho_{\gamma_u}\right)^{5/3}\leq Ch^{-5}\ell_0^{1/2}.
\end{equation}
{F}rom the Lieb-Thirring inequality (\ref{LT}) and the estimate
(\ref{eq:tfdf}) with $\alpha=0$, we conclude that 
$$
0\geq\Tr\left(\phi_u\left(-h^2\Delta-V^{\rm
      TF}\right)\phi_u\gamma_u\right) \geq
\mfr{1}{2}\Tr\left(-h^2\Delta(\phi_u\gamma_u\phi_u)\right)-Ch^{-3}\ell_0^{1/2},
$$
where we have used that $d(u)\leq 2\ell_0$ implies that $\ell(u)\leq
C\ell_0$. 
The density of the operator $\phi_u\gamma_u\phi_u$ is 
$\phi_u^2\rho_{\gamma_u}$. Thus,  
using the Lieb-Thirring inequality in the formulation
(\ref{LTdensity}) we arrive at (\ref{eq:rhogammu5/3}).

Using (\ref{eq:rhogammu5/3}), H\"older's inequality, the support property for
$\phi_u$, (\ref{eq:elldcom}), and the second property in (\ref{eq:phiuprop}) 
we arrive at the estimates
\begin{equation}\label{eq:hydL1L5/3}
 \|\phi_u^2\rho_{\gamma_u}\|_1\leq
 Ch^{-3}\ell(u)^{3/2}\quad\hbox{and}\quad
 \|\phi_u^2\rho_{\gamma_u}\|_{6/5}\leq
 Ch^{-3}\ell(u),
\end{equation}
for $u$ with $d(u)\leq 2\ell_0$. For these $u$ we also have
\begin{equation}\label{eq:TFL1L5/3}
 \|\phi_u^2(V^{\rm TF})^{3/2}\|_1\leq C\ell_0^{3/2}\quad\hbox{and}\quad
 \|\phi_u^2(V^{\rm TF})^{3/2}\|_{6/5}\leq C\ell_0,
\end{equation}
where we have used that from (\ref{eq:tfdf}) with $\alpha=0$, 
$V^{\rm TF}(x)\leq C d(x)^{-1}$.

We are now ready to prove that the density matrix $\gamma$ has the
stated properties. We begin with proving (\ref{eq:maingamma3}).
The density of $\gamma$ is 
$$
\rho_{\gamma}(x)=\int_{d(u)<R}\phi_u(x)^2\rho_{\gamma_u}(x)\ell(u)^{-3}du.
$$
{F}rom (\ref{eq:rhogammaproplf2}) we see that for $d(u)>2\ell_0$
we have 
$$
 \int\phi_u(x)^2\rho_{\gamma_u}(x)dx\leq \frac{1}{6\pi^2h^3}\int
 \phi_u(x)^2V^{\rm TF}(x)^{3/2}dx+C h^{-2+1/5}f(u)^{9/5}\ell(u)^{9/5}
$$
and from (\ref{eq:hydL1L5/3}) and (\ref{eq:TFL1L5/3}) we get for 
$d(u)\leq 2\ell_0$
that
$$
\int \phi_u(x)^2\rho_{\gamma_u}(x)dx\leq\frac{1}{6\pi^2h^3}\int
 \phi_u(x)^2V^{\rm TF}(x)^{3/2}dx+Ch^{-3}\ell_0^{3/2}.
$$
Hence using the first property of $\phi_u$ in (\ref{eq:phiuprop}) we
obtain
\begin{eqnarray*}
  \int\rho_{\gamma}(x)dx&\leq& \frac{1}{6\pi^2h^3}\int
  V^{\rm TF}(x)^{3/2}dx+
  Ch^{-3}\int_{d(u)\leq2\ell_0}\ell_0^{3/2}\ell(u)^{-3}du\\&&
  +C\int_{2\ell_0<d(u)<R}h^{-2+1/5}f(u)^{9/5}\ell(u)^{-6/5}du
  \\
  &\leq&\frac{1}{6\pi^2h^3}\int V^{\rm TF}(x)^{3/2}dx+
  Ch^{-3}\ell_0^{3/2}\\
  &&+Ch^{-2+1/5}\int_{2\ell_0<d(u)<R}\min\{d(u)^{-21/10},d(u)^{-18/5}\}du
  \\
  &\leq&\frac{1}{6\pi^2h^3}\int V^{\rm
    TF}(x)^{3/2}dx+Ch^{-2+1/5},
\end{eqnarray*}
where we have inserted the choice $\ell_0=h$, used that $h$ is small, and
controlled the integral over the region $2\ell_0<d(u)<R$  in
a way similar to the integral in (\ref{eq:DDsccom}), using 
the properties (\ref{fdefinition}) and (\ref{eq:elldcom2}) 
of $f$ and $\ell$.

We now come to the proof of (\ref{eq:maingamma2}). If we use the
Hardy-Littlewood-Sobolev inequality (\ref{Hardy-Littlewood-Sobolev}) 
we see that it is enough to estimate the $6/5$ norm 
\begin{eqnarray*} 
  \left\|\rho_\gamma-\frac{1}{6\pi^2h^3}(V^{\rm TF})^{3/2}\right\|_{6/5}
  &\leq&\int_{d(u)<R}\left\|\phi_u^2\left(\rho_{\gamma_u}
      -\frac{1}{6\pi^2h^3}(V^{\rm
        TF})^{3/2}\right)\right\|_{6/5}
  \ell(u)^{-3}du\\
  &&{}+\int_{d(u)>R}\left\|\frac{1}{6\pi^2h^3}(V^{\rm
        TF})^{3/2}\phi_u^2\right\|_{6/5}\ell(u)^{-3}du.
\end{eqnarray*}
If we use (\ref{eq:hydL1L5/3}) and (\ref{eq:TFL1L5/3}) when
$u$ with $d(u)\leq 2\ell_0$, (\ref{eq:rhogammaproplf1}) when 
$2\ell_0<d(u)<R$, and (\ref{eq:tflf}) when $d(u)>R$ we obtain   
\begin{eqnarray*}
  \left\|\rho_\gamma-\frac{1}{6\pi^2h^3}(V^{\rm TF})^{3/2}\right\|_{6/5}&\leq&
  Ch^{-3}\ell_0
  +Ch^{-21/10}\!\!\!\!\!\!\int\limits_{2\ell_0<d(u)<R}\!\!\!\!\!\!
  f(u)^{21/10}\ell(u)^{-7/5}du\\
  &&+Ch^{-3}\int_{d(u)>R}f(u)^3\ell(u)^{-1/2}du.
\end{eqnarray*}
Using as before the properties (\ref{fdefinition}) and
(\ref{eq:elldcom2}) we see that the first integral above is bounded by
a constant and the second integral (since we may assume that $R>1$) 
is bounded by
$R^{-3}$. Thus using the 
Hardy-Littlewood-Sobolev inequality (\ref{Hardy-Littlewood-Sobolev}) 
we obtain
$$
D\left(\rho_\gamma-\frac{1}{6\pi^2h^3}(V^{\rm
    TF})^{3/2}\right)\leq C\left\|\rho_\gamma-\frac{1}{6\pi^2h^3}(V^{\rm
    TF})^{3/2}\right\|_{6/5}^2\leq C(h^{-5+4/5}+h^{-6}R^{-6}).
$$

Finally we turn to proving (\ref{eq:maingamma1}). 
{F}rom the definition of $\gamma$, (\ref{eq:phiHphilf}) and
(\ref{eq:gammaproplf}) we obtain
\begin{eqnarray}
  \Tr \left[(-h^2\Delta - V^{\rm TF})\gamma\right]
    &=&\int_{d(u)<R}\Tr\left[\phi_u\gamma_u\phi_u
      \left(-h^2\Delta-V^{\rm TF}\right)\right] \ell(u)^{-3}du\nonumber\\
    &\leq&\int_{d(u)<R}\Tr\left[\phi_u
      \left(-h^2\Delta-V^{\rm TF}\right)\phi_u\right]_- \ell(u)^{-3}du\nonumber
    \\&&{}+Ch^{-2+1/5}\int_{2\ell_0<d(u)<R}f(u)^{19/5}\ell(u)^{-6/5}du.
    \label{eq:Hgamma1}
\end{eqnarray}
On the other hand, from (\ref{eq:tfphilow}) (used with $2R+1$ replaced
by $R$) we get
\begin{eqnarray}
  \Tr[-h^2\Delta-V^{\rm TF}]_-&\geq&
  \int_{d(u)<R}\Tr[\phi_u\left(-h^2\Delta-V^{\rm
      TF}-Ch^2\ell(u)^{-2}\right)\phi_u]_-  
  \ell(u)^{-3}du\nonumber\\&&-C(h^2R^{-2}+h^{-3}R^{-7}).\label{eq:Hgamma2}
\end{eqnarray}
Appealing to the Lieb-Thirring inequality (\ref{LT}) we see that for
all $0<\e<1$ we have
\begin{eqnarray*}
  \Tr[\phi_u\left(-h^2\Delta-V^{\rm
      TF}-Ch^2\ell(u)^{-2}\right)\phi_u]_- &\geq&
  (1-\e)\Tr[\phi_u\left(-h^2\Delta-V^{\rm TF}\right)\phi_u]_-\\&&{}
  -C\e^{-3/2}h^2\ell(u)^{-2}
  \\&&-C\e h^{-3}\int_{|x-u|<\ell(u)} 
  (V^{\rm TF}(x))^{5/2}dx. 
\end{eqnarray*}
Using (\ref{eq:tfdf}) and (\ref{eq:tflf}) both with $\alpha=0$ we find that 
$$
\int_{|x-u|<\ell(u)} 
V^{\rm TF}(x)^{5/2}dx\leq\left\{
  \begin{array}{ll}
    C\ell_0^{1/2},&\hbox{if } d(u)\leq 2\ell_0\\
    Cf(u)^5\ell(u)^3,&\hbox{if } d(u)> 2\ell_0
  \end{array}
\right..
$$
If $d(u)\leq2\ell_0$ we choose $\e= c\ell_0^{-1} h^2$ (as we did just 
before (\ref{eq:D-Dusmall})) and we get 
\begin{eqnarray}
\Tr[\phi_u\left(-h^2\Delta-V^{\rm
      TF}-Ch^2\ell(u)^{-2}\right)\phi_u]_- &\geq&
  \Tr[\phi_u\left(-h^2\Delta-V^{\rm TF}\right)\phi_u]_-\nonumber \\&&{}
  -Ch^{-1}\ell_0^{-1/2}\label{eq:Hgammausmall}
\end{eqnarray}
If $d(u)>2\ell_0$ we choose $\e= h^2\ell(u)^{-2}f(u)^{-2}$ 
(as we did just after (\ref{eq:D+lowerfinal})) 
and we get
\begin{eqnarray}
\Tr[\phi_u\left(-h^2\Delta-V^{\rm
      TF}-Ch^2\ell(u)^{-2}\right)\phi_u]_- &\geq&
  \Tr[\phi_u\left(-h^2\Delta-V^{\rm TF}\right)\phi_u]_-\nonumber \\&&{}
  -Ch^{-1}f(u)^3\ell(u)\label{eq:Hgammaularge}
\end{eqnarray}
Thus from (\ref{eq:Hgamma1}), (\ref{eq:Hgamma2}),
(\ref{eq:Hgammausmall}), and  (\ref{eq:Hgammaularge}) we get
\begin{eqnarray*}
  \Tr \left[(-h^2\Delta - V^{\rm TF})\gamma\right]&\leq&
  \Tr[-h^2\Delta-V^{\rm TF}]_-
  +C(h^{-1}\ell_0^{-1/2}+h^2R^{-2}+h^{-3}R^{-7})\\&&
  +C\int\limits_{d(u)<R}h^{-2+1/5}f(u)^{19/5}\ell(u)^{-6/5}
  +h^{-1}f(u)^3\ell(u)^{-2}du.
\end{eqnarray*}
The estimate (\ref{eq:maingamma1}) now follows from a calculation
almost identical to the one given right after (\ref{eq:DDsccom}).

\end{proof}

\section{Proof of the Scott correction for the molecular ground state
  energy}\label{sec:scott}

The proof of the Scott correction Theorem~\ref{main theorem} is now a
fairly standard application of the results in the previous sections. 
We begin with giving the proof of the lower bound. 

\begin{lemma}[Lower bound] \label{lower bound} Let $R$ and $Z$ be as in the
statement of Theorem~\ref{main theorem}. 
Then, the ground state energy for a neutral molecule satisfies
$$ E(Z,R) \ge E^{\rm TF}(Z,R)  + \mfr{1}{2} \sum_j Z_j^2 + 
    {\cal O} (|Z|^{2-1/30}) .
$$
\end{lemma}

\begin{proof} The starting point is the Lieb-Oxford inequality 
  (\ref{Lieb-Oxford}), from which we conclude that if $\psi$ is a
  $Z$-particle ($N=Z$) wave function we have
$$
\langle \psi, H(Z,R)\psi\rangle
\ge \sum_{i=1}^Z \left\langle\psi,\big[-\mfr{1}{2}\D_i -
  V(Z,R,x_i)\big] 
  \psi\right\rangle
+ D(\rho_\psi) - C\int \rho_\psi^{4/3}.
$$

In order to bound the last term we use the 
many-body version of the Lieb-Thirring inequality (\ref{eq:LTmbcase}). 
For all $0<\e<1/2$ we
have 
\begin{eqnarray*} 
  \Big\langle\psi,\e\sum_{i=1}^Z-\mfr{1}{2}\D_i \psi\Big\rangle 
  - C\int \rho_\psi^{4/3}
  &\ge&\e\int\rho_\psi^{5/3} - C \int\rho_\psi^{4/3}
  \\
  &\ge&\e\int\rho_\psi^{5/3} - C \left(\int\rho_\psi^{5/3}\right)^{1/2}
  \left(\int\rho_\psi\right)^{1/2}
  \\
  &\ge&-\e^{-1} C \int \rho_\psi =-C\e^{-1}Z.
\end{eqnarray*}
Here we have used H\"older's inequality for the $\rho^{4/3}$ integral
and used that $\psi$ is a $Z$-particle state.
Thus 
\begin{eqnarray*}
  \langle \psi, H(Z,R)\psi\rangle&\ge& \left\langle \psi, 
    \sum_{i=1}^Z (-(1-\e)\mfr{1}{2}\D_i - V(Z,R,x_i)) \psi\right\rangle+
  D(\rho_\psi)
  - C\e^{-1}Z\\
  &\ge& \left\langle \psi, 
    \sum_{i=1}^Z (-(1-\e)\mfr{1}{2}\D_i - \V(Z,R,x_i))
    \psi\right\rangle+
  D(\rho-\rho^{\rm TF}(Z,R,\cdot))\\&& - D(\rho^{\rm TF}(Z,R,\cdot))- C\e^{-1}Z\\
  &\geq&
 2\,\Tr\big[-\mfr{1}{2}(1-\e)\D - \V(Z,R,\cdot)\big]_- 
   - D(\rho^{\rm TF}(Z,R,\cdot))- C\e^{-1}Z.
\end{eqnarray*}
Here we have applied (\ref{eq:tfpotgeneral}), the fact that the Coulomb
kernel is positive definite such that $D(\rho-\rho^{\rm TF})\geq0$, and the
Fermionic property of the wave function.

If we now use the scaling property (\ref{scaling}) we find that 
$$
\Tr\big[-\mfr{1}{2}(1-\e)\D - \V(Z,R,\cdot)\big]_- 
=|Z|^{4/3}\Tr\big[-\mfr{1}{2}(1-\e)|Z|^{-2/3}\D - 
\V({\bf z},{\bf r},\cdot)\big]_-,
$$
where ${\bf z}=(z_1,\ldots,z_M)$ and ${\bf r}=(r_1,\ldots,r_M)$. 
Using now (\ref{eq:main1}) (with $h=\sqrt{(1-\e)/2}|Z|^{-1/3}$)
and (\ref{eq:sc=tf}) we see that
\begin{eqnarray*}
 2 \Tr\big[-\mfr{1}{2}(1-\e)\D - \V(Z,R,\cdot)\big]_- 
  &=&(1-\e)^{-3/2}|Z|^{7/3}\left(E^{\rm TF}({\bf
      z},{\bf r})+D(\rho^{\rm TF}({\bf
      z},{\bf r},\cdot)\right)\\&&
  +(1-\e)^{-1}\frac{|Z|^2}{2}\sum_{k=1}^Mz_k^2+O(|Z|^{2-1/30})\\
  &=&(1-\e)^{-3/2}\left(E^{\rm TF}(Z,R)+D(\rho^{\rm TF}(Z,R,\cdot)\right)\\&&
  +\mfr{1}{2}(1-\e)^{-1}\sum_{k=1}^MZ_k^2+O(|Z|^{2-1/30}).
\end{eqnarray*}
We have here used the TF scaling $E^{\rm TF}(Z,R)=|Z|^{7/3}E^{\rm
  TF}({\bf z},{\bf r})$ and $D(\rho^{\rm
  TF}\left(Z,R,\cdot\right)=|Z|^{7/3}D\left(\rho^{\rm TF}({\bf z},{\bf
    r},\cdot)\right)$.  Choosing $\e=|Z|^{-2/3}$ completes the proof
of the lemma.
\end{proof}

\begin{lemma}[Upper bound] \label{upper bound} Let $R$ and $Z$ 
satisfy the conditions from Theorem~\ref{main theorem}. 
Then, the ground state energy for a neutral
molecule satisfies 
$$ E(Z,R) \le E^{\rm TF} (Z,R) + \mfr{1}{2} \sum_j Z_j^2 + 
    {\cal O} (|Z|^{2-1/30}) .
$$
\end{lemma}

\begin{proof} The starting point now is Lieb's variational principle, 
Theorem~\ref{Lieb's Variational Principle}. 
By a simple rescaling the variational principle states that for any
density matrix $\gamma$ on $L^2(\R^3)$ with $2\Tr\gamma\leq Z$ we have  
$$
  E(Z,R)\leq|Z|^{4/3}\left(2\Tr\left[\left(-\mfr{1}{2}|Z|^{-2/3}\Delta
      -V({\bf z},{\bf r},x)\right)\gamma\right]
  +|Z|D(2|Z|^{-1}\rho_\gamma)\right).
$$
As for the lower bound we bring the TF-potential 
into play
\begin{eqnarray}
  |Z|^{-4/3}E(Z,R)&\leq&2\Tr\left[\left(-\mfr{1}{2}|Z|^{-2/3}\Delta
      -V({\bf z},{\bf r},x)\right)\gamma\right]
  +|Z|D(2|Z|^{-1}\rho_\gamma)\nonumber\\
  &=&
  2\Tr\left[\left(-\mfr{1}{2}|Z|^{-2/3}\Delta
      -\V({\bf z},{\bf r},x)\right)\gamma\right]\nonumber\\&&
  +|Z|D\left(2|Z|^{-1}\rho_\gamma-\rho^{\rm TF}({\bf z},{\bf r},\cdot)\right)
  -|Z|D(\rho^{\rm TF}({\bf z},{\bf r},\cdot)).\label{eq:upper1}
\end{eqnarray}
We now choose a density matrix $\widetilde{\gamma}$ 
according to Theorem~\ref{TF} with $h=\sqrt{1/2}|Z|^{-1/3}$. 
Note that with this choice of $h$ we have that 
$$
(6\pi^2h^3)^{-1}\V({\bf z},{\bf r},x)^{3/2}=|Z|\rho^{\rm TF}({\bf
  z},{\bf r},x)/2.
$$
Since $\int\rho^{\rm TF}({\bf z},{\bf
  r},x)=\sum_{j=1}^Mz_j=1$ we see from (\ref{eq:maingamma3}) that
$$
2\Tr\widetilde\gamma\leq |Z|+C|Z|^{2/3-1/15}=|Z|(1+C|Z|^{-1/3-1/15}).
$$
Thus if we define $\gamma=(1+C|Z|^{-1/3-1/15})^{-1}\widetilde{\gamma}$ 
we see that the condition $2\Tr\gamma\leq |Z|$ is satisfied.

Using (\ref{eq:maingamma2}) we see that 
$$
|Z|D\left(2|Z|^{-1}\rho_{\widetilde\gamma}-\rho^{\rm TF}({\bf z},{\bf
    r},\cdot)\right) \leq C|Z|^{2/3-4/15} ,
$$
and thus 
\begin{eqnarray}
  \lefteqn{|Z|D\left(2|Z|^{-1}\rho_\gamma-\rho^{\rm TF}({\bf z},{\bf
        r},\cdot)\right)}&&\nonumber\\ &\leq&
  C|Z|(1+C|Z|^{-1/3-1/15})^{-2}
  D\left(2|Z|^{-1}\rho_{\widetilde\gamma}-\rho^{\rm TF}({\bf z},{\bf
      r},\cdot)\right)\nonumber\\&&+
  C|Z|^{1/3-2/15}D\left(\rho^{\rm TF}({\bf z},{\bf
      r},\cdot)\right)\leq C |Z|^{2/3-4/15},\label{eq:upper2}
\end{eqnarray}
where we have used the triangle inequality for $\sqrt{D}$, and
that $D\left(\rho^{\rm TF}({\bf z},{\bf r},\cdot)\right)\leq C$.

Finally, if we use (\ref{eq:main1}), (\ref{eq:maingamma1}), and
(\ref{eq:sc=tf})
we arrive at 
\begin{eqnarray*}
  2\Tr\left[\left(-\mfr{1}{2}|Z|^{-2/3}\Delta
      -\V({\bf z},{\bf r},x)\right)\widetilde\gamma\right]&\leq&
  |Z|\left(E^{\rm TF}({\bf
      z},{\bf r})+D(\rho^{\rm TF}({\bf
      z},{\bf r},\cdot)\right)\\&&
  +\frac{|Z|^{2/3}}{2}\sum_{k=1}^Mz_k^2+O(|Z|^{2/3-1/30}).
\end{eqnarray*}
Since $E^{\rm TF}({\bf z},{\bf r})\leq C$ and 
$D\left(\rho^{\rm TF}({\bf z},{\bf r},\cdot)\right)\leq C$ we see that the
same estimate holds for $\widetilde\gamma$ replaced by $\gamma$. 
If we insert this estimate together with (\ref{eq:upper2}) into
(\ref{eq:upper1}) and use again that 
$E^{\rm TF}(Z,R)=|Z|^{7/3}E^{\rm TF}({\bf z},{\bf r})$ we arrive at
the upper bound in the lemma.
\end{proof}

\appendix
\section{Appendix: Results on the new coherent states}

Before we prove the trace formula (\ref{trace formula}) and the representation 
(\ref{thm:coherentrepresentation}) we need some simple lemmas. The
proof of the first one is a straightforward calculation which it is left
to the reader to check.

\begin{lemma}\label{lm:Gkernel} Let ${\cal G}_{u,q}$ be defined as in (\ref{new coherent
    states}), then its integral kernel is
  \begin{equation}{\cal G}_{u,q}(x,y) = (\pi h)^{-n/2} e^{-a\left(\frac{x+y}{2}-u\right)^2
      +iq(x-y)/h -\frac{1}{4h^2a}(x-y)^2} . \label{eq:Gkernel}
  \end{equation}
\end{lemma}
\begin{lemma} \label{lm:fkernel}
  Let $B_0\in\mathbb R,B_{1,2}\in\mathbb R^n$ and $\hat{A} = 
  B_0 + B_1 {\x} - ihB_2 \nabla$ be a linear 
  combination of the identity, multiplication and momentum operator.
  Let $f$ be a polynomially bounded, measurable function on $\mathbb
  R^n$ with values in $\mathbb R$.  Then,
  $$f(B_0+B_1{\x}-ihB_2\nabla)(x,y) = \int \, 
  f(B_0+B_1\left(\mfr{x+y}{2}\right)+B_2p) e^{ip(x-y)/h} \frac{dp}{(2\pi h)^n} 
  $$
  as a distributional kernel.
\end{lemma}
\begin{proof} 
  First assume that $B_1$ and $B_2$ are not orthogonal. By applying
  the unitary transformation $(U\psi)(x) = \exp{[-\frac{i}{2hB_1\cdot
      B_2}(B_1 x)^2]}\psi(x)$ and the Spectral Theorem we get $U^{-1}
  \,f(B_0 + B_1{\x}-ihB_2\nabla)\, U = f(B_0 -ih B_2\nabla) $.  Hence,
  \begin{eqnarray*} \lefteqn{f(B_0+B_1{\x} - ihB_2\nabla)(x,y)}
    \\
    &=&\int f(B_0+B_2p) \,e^{-\frac{i}{2hB_1\cdot B_2}((B_1x)^2-(B_1y)^2)+ip(x-y)/h}
    \, \frac{dp}{(2\pi h)^n}
    \\
    &=&\int f(B_0 + B_1 \left(\mfr{x+y}{2}\right) + B_2 p)\,e^{ip(x-y)/h}
    \, \frac{dp}{(2\pi h)^n} .
  \end{eqnarray*}      
  The case $B_1\cdot B_2=0$ follows then, say by continuity.
\end{proof}

\begin{proof}[Proof of Theorem~\ref{thm:traceidentity}]
The proof is actually a  fairly standard exercise in calculations with Fourier
integrals, but we shall do it here carefully. 

Let us for simplicity  call ${\cal G}={\cal G}_{u,q}$.
Since $f$ is polynomially bounded we have that 
$(1+\hat{A}^{2N})^{-1}f(\hat{A})$ extends to a bounded operator 
when $N$ is a large enough integer. {F}rom the explicit expression 
(\ref{eq:Gkernel}) for the
integral kernel of ${\cal G}$ we immediately see that 
$(1+\hat{A}^{2N}){\cal G}(x,y)$ is in $L^2(\R^n\times\R^n)$. Thus, the 
operator $(1+\hat{A}^{2N}){\cal G}$ is Hilbert-Schmidt. It follows
that $f(\hat{A}){\cal G}$ is Hilbert-Schmidt.
Likewise $(1+|x|)^{-M}V(x)$ is bounded for $M$ large enough. 
and thus $\cG V(\x)$ extends to a  Hilbert-Schmidt operator.  
Moreover if we define 
$$
  f_\varepsilon(s)=f(s)e^{-\varepsilon s^2}\hbox{ and }
  V_\varepsilon(x)=V(x)e^{-\varepsilon x^2} ,
$$
we have 
$$
\lim_{\varepsilon\to0}\Tr\left[{\cal G}f_\varepsilon(\hat{A}){\cal G}V_\varepsilon\right]
=\Tr\left[{\cal G}f(\hat{A}){\cal G}V\right].
$$
The trace on the left can be immediately calculated from
Lemmas~\ref{lm:Gkernel} and \ref{lm:fkernel}.
\end{proof}

\begin{proof}[Proof of Theorem~\ref{trace formula}]
We proceed as in the previous proof. We consider first the case $V=0$.
We have that $(1-h^2\Delta)^{-1}F(-ih\nabla)$ 
is a bounded operator and thus  
$\phi F(-ih\nabla)\phi\cG$ is a Hilbert-Schmidt operator.
It follows moreover that if we define
$$
  f_\varepsilon(s)=f(s)e^{-\varepsilon s^2}\hbox{ and }
  F_\varepsilon(p)=F(p)e^{-\varepsilon p^2} ,
$$
then the operators 
$$
f_\varepsilon(\hat{A}){\cal G}\hbox{ and }
\phi F_\varepsilon(-ih\nabla)\phi\cG
$$
converge in Hilbert-Schmidt norm as $\varepsilon\to0$.
We  therefore have
$$ 
\mbox{Tr}\big[{\cal G}\,f(\hat{A}) \,{\cal G}\,
\phi F(-ih\nabla)\phi \big]=
\lim_{\varepsilon\to0}\lim_{\delta\to0}\mbox{Tr}\big[{\cal G}\,f_\delta(\hat{A}) \,{\cal G}\,
\phi F_\varepsilon(-ih\nabla)\phi \big].
$$
The trace on the right can be written as an absolutely convergent integral
\begin{eqnarray*}\lefteqn{(2\pi h)^{2n}
    \mbox{Tr}\big[{\cal G}\,f_\delta(\hat{A}) \,{\cal G}\,
    \phi F_\varepsilon(-ih\nabla)\phi \big]
    }\\
  &=&\int\,f_\delta(B_0 + B_1\mfr{x+y}{2} + B_2p) 
  e^{ip(x-y)/h +i\eta(z-z')/h}{\cal G}_{u,q}(y,z){\cal G}_{u,q}(z'x)
  \\
  &&\times\,\phi(z)\phi(z')F_\varepsilon(\eta)
  \, d\eta dx dy dz dz'dp 
\end{eqnarray*}
We now perform the integration in the variable $x-y$ and we rename the 
integration variable $(x+y)/2$ as $v$. We subsequently change variables so that 
$(z,z',\eta)$ are replaced by
$(z+h^2ab(u-v)+v,z'+h^2ab(u-v)+v,\eta+p+h^2(ab)(q-p))$.
This eventually gives 
\begin{eqnarray}
  \lefteqn{(2\pi h)^{2n} (\pi/4b)^{n/2}\mbox{Tr}\big[{\cal G}\,f_\delta(\hat{A}) 
    \,{\cal G}\,\phi F_\varepsilon(-ih\nabla)\phi \big]
    }\nonumber\\
  &=&\int f_\delta(B_0+B_1v+B_2p)
  F_\varepsilon(\eta+w(p,q))e^{-b(p-q)^2-b(u-v)^2} 
  \label{eq:5fourier}\\
  &&\times\, \phi(z+w(v,u)) \phi(z'+w(v,u))
  e^{i\eta(z-z')/h}
  e^{-b\left(\frac{z-z'}{2}\right)^2 -\frac{1}{4h^2b} (z+z')^2}\, 
  dz dz'd\eta dvdp,\nonumber
\end{eqnarray}
where to avoid lengthy expressions we have introduced the function 
$w(s,t)= s+h^2ab(t-s)$ for $s,t\in\R^n$.
We now expand $F(w(p,q)+\eta)$ around $w(p,q)$. I.e., we
write
\begin{eqnarray*}
  F_\varepsilon(w(p,q)+\eta)=\Bigl[F(w(p,q))+\eta\cdot\nabla
  F(w(p,q))+R_F(w(p,q),\eta)\Bigr] 
  e^{-\varepsilon(w(p,q)+\eta)^2},
\end{eqnarray*}
where according to the assumption that all second derivatives of $F$
are bounded we have that the remainder term satisfies
$|R_F(p+h^2ab(q-p),\eta)|\leq C \eta^2$, with a constant depending on the
bound on the second derivatives of $F$. We shall estimate the error
coming from the remainder term below, but we first consider the
contribution from the two main terms.
If we insert the two main terms in the expansion into the integral
(\ref{eq:5fourier}) above we see that they
give integrals in which the $\eta$ integration can be
performed explicitly. After letting $\delta,\varepsilon\to0$ we obtain for
these leading terms
\begin{equation}
\begin{array}{rl}  
(2\pi h)^{n}\displaystyle\int& f(B_0+B_1v+B_2p)
  F(w(p,q))e^{-b(p-q)^2-b(u-v)^2} \\
  &\phi(z+w(v,u))^2
  e^{-\frac{1}{h^2b} z^2}\, dzdvdp.
\end{array}\label{eq:aftereta}
\end{equation}
Here we may now expand the function $\phi^2$ 
$$
  \phi(z+w(v,u))^2=\phi(w(v,u))^2+z\cdot\nabla\phi(w(v,u))^2
  +R_{\phi}(w(v,u),z),
$$
where $|R_{\phi}(h^2ab(u-v)+v,z)|\leq Cz^2$ with a constant that
depends on the bound on $\phi$, and its first and second derivatives.
Hence the integral in (\ref{eq:aftereta}) may be written as
$$\begin{array}{rl}\displaystyle
  (2\pi h)^{n}(b/4\pi)^{n/2}\int&f(B_0+B_1v+B_2p)
  e^{-b(p-q)^2-b(u-v)^2} F(w(p,q))\\ & \displaystyle\times 
  \left[\phi(w(v,u))^2 + E_1(u,v)\right]dvdp,
\end{array}
$$
where $|E_1(u,v)|\leq Ch^2b$.

We now return to estimating the contribution from the remainder term 
$R_F$. Note that for all
integers $1\leq k\leq (n/2)+2$ we have 
\begin{eqnarray*}
\lefteqn{(1+\eta^2/(bh^2))^{k}\left|\int\phi(z+w(v,u))
  \psi(z'+w(v,u))
  e^{i\eta(z-z')/h}
  e^{-b\left(\frac{z-z'}{2}\right)^2}\, d(z-z')\right|}&&
\\ &\leq&C\int\left|(1-b^{-1}\Delta_{z-z'})^{k}\left(
  \phi(z+w(v,u))\phi(z'+w(v,u))
  e^{-b\left(\frac{z-z'}{2}\right)^2}\right)\right|d(z-z')\\
&\leq&Cb^{-n/2},
\end{eqnarray*}
where we have used that $b>1$. Here the constant depends on
the bound on $\phi$, and its first $n+4$ derivatives.
Thus the relevant contribution to the integral (\ref{eq:5fourier}) coming from
the error term $R_F$ can be estimated by 
\begin{eqnarray*}
&&\Biggl|\int R_F(p+h^2ab(q-p),\eta) \phi(z+w(v,u))
  \phi(z'+w(v,u))
  e^{i\eta(z-z')/h}\\&&\times\,
  e^{-b\left(\frac{z-z'}{2}\right)^2 -\frac{1}{4h^2b} (z+z')^2}\, 
  dz dz'd\eta\Biggr|\\
&&\leq Cb^{-n/2}
  \int(1+\eta^2/(bh^2))^{-k}\eta^2e^{-\frac{1}{h^2b} \tilde{z}^2}\, 
  d\tilde{z}d\eta 
 \leq Cb^{n/2}h^{2n} bh^2,
\end{eqnarray*}
where we have chosen $k$ so as to make the integral finite. We can
always do this without violating $1\leq k\leq (n/2)+2$.
Thus after taking the limit $\delta\to0$ we obtain the statement of
the theorem. 

The case when $F=0$ and $V\ne0$ is similar but much simpler since
we may start with Theorem~\ref{thm:traceidentity} and expand $V$.
\end{proof}

\begin{proof}[Proof of theorem \ref{thm:coherentrepresentation}]
Since $\s$ is a sum of a function of $q$ and a function of $u$ it is
enough to consider only one of the terms, say, $V$. 
Let as before $G_b(x)=(b/\pi)^{n/2}e^{-b x^2}$. It follows immediately from
(\ref{eq:Gkernel}) that
\begin{eqnarray}
  \int {\cal G}_{u,q}^2\frac{dq}{(2\pi h)^n}&=&G_b({\x}-u),\nonumber\\
  \int {\cal G}_{u,q}({\x}-u){\cal G}_{u,q}\frac{dq}{(2\pi h)^n}
  &=&(1-h^2ab)({\x}-u)G_b({\x}-u).\label{eq:firstorder}
\end{eqnarray}
As a consequence we have
\begin{eqnarray*}
  \lefteqn{V({\x}) -\int {\cal G}_{u,q}
    \Bigl(V(u)+\mfr{1}{4b}\Delta V(u)+\nabla V(u)
      \cdot ({\x}-u)
    \Bigr){\cal G}_{u,q}\frac{du dq}{(2\pi h)^n}}\hspace{4truecm}&&
  \\&=&
  \int G_b({\x}-u)\Bigl(V({\x})-\Bigl(V(u)+\mfr{1}{4b}\Delta V(u)
  \\&&{}+(1-h^2ab)\nabla V(u) \cdot ({\x}-u)\Bigr)
  \Bigr)du.
\end{eqnarray*}
Using Taylors' formula we have 
$$
V({x})=V(u)+\nabla V(u)\cdot ({x}-u)
+{\textstyle\frac{1}{2}}\sum_{ij}\partial_i\partial_j
V({x})({x}_i-u_i)({x}_j-u_j)-{\cal
  R}_1({x},u)
$$
where 
$$
{\cal R}_1({x},u)={\textstyle\frac{1}{2}}\int_0^1\sum_{i,j,k}
\partial_i\partial_j\partial_kV(u+t(x-u))
({x}_i-u_i)({x}_j-u_j)({x}_k-u_k)(1-(1-t)^2)dt.
$$
Since $\int x_ix_jG_b(x)dx=\frac{1}{2b}\delta_{ij}$ we have
\begin{eqnarray*}
  \lefteqn{V({\x}) -\int {\cal G}_{u,q}
  \left(V(u)+\mfr{1}{4b}\Delta V(u)+\nabla V(u)
    \cdot ({\x}-u)
  \right){\cal G}_{u,q}\frac{du dq}{(2\pi h)^n}}&&
  \\&=&\int G_b({\x}-u)
  \Bigl(\mfr{1}{4b}\left(\Delta V({\x})-\Delta V(u) \right)
  +h^2ab\nabla V(u)\cdot ({\x}-u)-{\cal R}_1({\x},u)
  \Bigr)du\\
  &=&\int G_b({\x}-u)
  \left(\mfr{1}{4b}\left(\Delta V({\x})-\Delta V(u) \right)
    -\mfr{1}{2}h^2a\Delta V(u)-{\cal R}_1({\x},u)
  \right)du
\end{eqnarray*}
where the last identity follows by integration by parts. 
The theorem now follows easily since $a\leq b$ and 
$$
\Delta V({x})-\Delta
V(u)=\int_0^1\sum_i\partial_i\Delta
V(u+t({x}-u))({x}_i-u_i)dt.
$$
\end{proof}

\section{Appendix: A localization theorem}

\begin{thm} \label{partition}
Consider $\phi\in C^\infty_0(\mathbb R^n)$ with support in the 
ball $\{|x|\leq1\}$ and satisfying $\int\phi^2(x)\,dx=1$. 
Assume that $\ell:\R^n\to \R$ is a $C^1$ map 
satisfying $0<\ell(u)\leq1$ and $\|\nabla\ell\|_\infty<1$. 
Let $J(x,u)$ be the Jacobian of the map $u\mapsto
\frac{x-u}{\ell(u)}$, i.e.
$$
J(x,u)=
\ell(u)^{-n}\left|\det\left[\frac{(x_i-u_i)\partial_j\ell(u)}{\ell(u)} 
      + \delta_{ij}\right]_{ij}\right|.
$$
We set $\phi_{u}(x):=\phi\Big(\frac{x-u}{\ell(u)}\Big)\sqrt{J(x,u)}
\ell(u)^{n/2}$.
Then, for all $x\in \R^n$ 
\begin{equation}\label{eq:philoc}
  \int_{\R^n} \phi_{u}^2(x)\ell(u)^{-n}\, du = 1 
\end{equation}
and for all multi-indices $\alpha$ we have
\begin{equation}\label{eq:phiuestimate}
  \|\p^\alpha\phi_{u}\|_\infty \leq \ell(u)^{-|\alpha|}
  C_\alpha\max_{|\beta|\leq|\alpha|}\|\p^\beta\phi\|_\infty,
\end{equation}
where $C_\alpha$ depends only on $\alpha$.
\end{thm}
\begin{proof}
  In order to prove (\ref{eq:philoc}) it is of course enough to
  consider the case $x=0$.  The identity follows from the change of
  variables formula if we can show that the map $F:\R^n\to\R^n$ given
  by $F(u)=-u/\ell(u)$ is a bijection of
  $F^{-1}\left(\{|x|\leq1\}\right)$ onto $\{|x|\leq1\}$.
  
  The map is, in fact, onto $\R^n$ since $F(0)=0$ and $|F(u)|\geq
  |u|$.  Hence for all $u\in \R^n$ there exists $t\in\R$ with $-1\leq
  t\leq 0$ such that $F(tu)=u$.
  
  That the map is also injective on $F^{-1}\left(\{|x|\leq1\}\right)$
  follows since for $u\ne0$ we may write $F(tu)=-g(t)u$ and the map
  $g:\mathbb R\to\mathbb R$ is monotone increasing for all $t$ for
  which $|g(t)||u|\leq1$.  In fact, $g(t) = t/\ell(tu)$ and thus
  \begin{eqnarray*}
    g'(t)&=&\ell(tu)^{-1}-t\ell(tu)^{-2}\nabla\ell(tu)\cdot u  
    =\ell(tu)^{-1}[1-t\ell(tu)^{-1}\nabla\ell(tu)\cdot u]\\
    &\geq& \ell(tu)^{-1}[ 1-\|\nabla\ell\|_\infty |g(t)||u|]>0.
  \end{eqnarray*}
  Note that
  $\phi_u(x)=\widetilde{\phi}_u\left(\frac{x-u}{\ell(u)}\right)$,
  where 
  $$
  \widetilde{\phi}_u(x)=\phi(x)\left|\det\left[x_i\p_j\ell(u)+\delta_{ij}
    \right]_{ij}\right|.
  $$
  The estimates (\ref{eq:phiuestimate}) follow since $
  \|\partial^\alpha\widetilde{\phi}_u\|_\infty\leq
  C\max_{|\beta|\leq|\alpha|}\|\p^\beta\phi\|_\infty.  $
\end{proof}

\end{document}